\documentclass[12pt]{article}
\usepackage[english]{babel}
\usepackage[utf8]{inputenc}
\usepackage[T1]{fontenc}
\usepackage{authblk}
\usepackage{hyperref}
\usepackage{caption}
\usepackage{graphicx}
\usepackage{amsmath,mathrsfs,dsfont}
\usepackage{amsthm}
\usepackage[all]{xy}
\usepackage{graphics}
\usepackage{tensor}
\usepackage{verbatim}
\usepackage{nicefrac}
\usepackage{amsfonts}
\usepackage{amssymb,latexsym}
\usepackage{color}
\usepackage{mathrsfs,dsfont}
\usepackage{appendix}
\usepackage{slashed}
\usepackage{cite}
\usepackage{comment}
\usepackage{bookmark}

\newcommand{\be}{\begin{equation}}
\newcommand{\ee}{\end{equation}}
\newcommand{\bea}{\begin{eqnarray}}
\newcommand{\eea}{\end{eqnarray}}
\newcommand{\nn}{\nonumber}

\numberwithin{equation}{section}

\definecolor{verde}{cmyk}{.83,.21,1,.08}

\def\con{\color{black}} 

\let\Re\relax
\DeclareMathOperator{\Re}{\mathfrak{Re}}
\DeclareMathOperator{\Tr}{Tr}

\begin{document}

\pdfbookmark{Spectral action approach to higher derivative gravity}{name1}

\title{Spectral action approach to higher derivative gravity}
\author[1]{Ravi Mistry\thanks{ravi.mistry.r@gmail.com}}
\author[1,2]{Aleksandr Pinzul\thanks{aleksandr.pinzul@gmail.com}}
\author[1,3]{Les{\l}aw Rachwa{\l}\thanks{grzerach@gmail.com}}
\affil[1]{Universidade de Bras\'{\i}lia, Instituto de F\'{\i}sica 70910-900, Bras\'{\i}lia, DF, Brazil}
\affil[2]{International Center of Physics C.P. 04667, Bras\'{\i}lia, DF, Brazil}
\affil[3]{Department of Physics, Faculty of Nuclear Sciences and Physical Engineering, Czech Technical University in Prague,
11519, Prague, \newline Czech Republic}
\date{}
\maketitle

\pdfbookmark[1]{Abstract}{name2}
\begin{abstract}
We study the spectral action approach to higher derivative gravity. The work focuses on the classical aspects. We derive the complete and simplified form of the purely gravitational action up to the 6-derivative terms. We also derive the equivalent forms of the action, which might prove useful in different applications, namely Riemann-- and Weyl--dominated representations. The spectral action provides a rather rigid structure of the higher derivative part of the theory. We discuss the possible consequences of this rigidness. As one of the applications, we check whether the conformal backgrounds are preferred in some way on the classical level, with the conclusion that at this level, there is no obvious reason for such a preference, the space $S^1 \times S^3$ studied in earlier works being a special case. Some other possible properties of the higher derivative gravity given by the spectral action are briefly discussed.
\end{abstract}

\hypersetup{bookmarksdepth=2}
\pdfbookmark[1]{1 Introduction}{name3}
\hypersetup{bookmarksdepth=-2}
\section{Introduction}
\label{sec1}
\hypersetup{bookmarksdepth}

In the absence of a completely satisfactory theory of Quantum Gravity (QG), it is important to look for some universal features of the future theory as well as to establish some connections between different approaches towards it. In this paper we look at possible benefits one can gain in combining two such seemingly unrelated approaches to (quantum) gravity as spectral action and higher derivative gravity.

The spectral action approach \cite{Chamseddine:1996rw,Chamseddine:1996zu,Chamseddine:2008zj} appeared as a natural but non-trivial development of the approach to Standard Model based on the methods of non-commutative geometry \cite{Connes:1990qp,Connes:1994yd}. Though this approach does not address directly the question of QG, one cannot say that it is purely classical as it is spectral by its nature. The main advantage of the spectral action formulation is, from our point of view, that it provides a \textit{geometric} unification of gravity with Standard Model: both, gravity and Yang-Mills (YM) sector, are treated as ``perturbations'' of some non-commutative geometry while the coupling to fermionic matter takes a form of the usual Dirac action. The full spectral action is (almost) completely fixed by the choice of the so-called spectral triple, $(\mathcal{A},\mathrm{D},\mathcal{H})$. Here $\mathcal{A}$ is an algebra of smooth functions on the generalized spacetime of the form $M\times F$, where $M$ is the usual classical smooth spacetime and $F$ is some finite matrix geometry; $\mathrm{D}$ is the relevant Dirac operator defined on the full non-commutative geometry; and $\mathcal{H}$ is a Hilbert space, on which all these objects are represented as operators satisfying some conditions (see \cite{Connes:1994yd,GraciaBondia:2001tr} for the details on non-commutative geometry and, e.g., \cite{vanSuijlekom:2015iaa} for the non-commutative geometry approach to Standard Model). Thus the full spectral action is given by \cite{Chamseddine:1996zu}
\be\label{spectralaction}
S_{{\rm spec}}=\Tr\chi\left(\frac{\mathrm{D}^2}{\Lambda^2}\right) + \langle\psi|\mathrm{D} |\psi\rangle,
\ee
where $\chi=\chi(p)$ is some positive cut-off function and $\Lambda$ is some characteristic energy scale.\footnote{The presence of this function $\chi(p)$ and the scale $\Lambda$ is the reason why the spectral triple \textit{almost} fixes the spectral action. As we will discuss, they should be considered as independent inputs of the theory, presumably fixed by some fundamental theory of QG.} While the Dirac-type second term of (\ref{spectralaction}) describes the matter-geometry coupling (including coupling to YM gauge fields) the first term is completely spectral and describes the dynamics of geometry, i.e. gravity plus Yang-Mills.\footnote{Another nice feature of the spectral action is that Higgs field finds its natural place on the geometry side of the picture (rather than on the matter side which is completely encoded in the Hilbert space $\mathcal{H}$).} In the case of pure gravity, i.e. using the usual Dirac operator, this term correctly ``predicts'' the standard Einstein-Hilbert (E-H) action with cosmological constant as the first two terms of some asymptotic expansion. The higher terms in this expansion produce action terms depending on higher orders of curvature. The beauty of the spectral action is that these terms depend only on three inputs: the choice of the Dirac operator $\mathrm{D}$, the scale $\Lambda$ and the cut-off function $\chi(p)$.\footnote{The cut-off function is a non-local object so, strictly speaking it introduces the infinite number of parameters, but in a very controlled way, see below.} Also, each level of the expansion is a very precise combination of terms with the same derivative order. E.g., as we will see, the next term in the expansion (after the one reproducing E-H action) has precisely four derivatives of the metric tensor. However, it is worth to emphasize that in principle the series for the spectral action continues up to infinity, so the full (re-summed) theory would contain infinitely many derivatives and therefore would be non-local. The appearance of these higher-order terms (starting with the ones quadratic in curvature) is what makes the direct connection with the higher derivative gravity theories. Such theories contain more (than two) derivatives on metric in their corresponding classical equations of motion (EOM). As we have discussed above the presence of higher derivative terms (in the form of terms with higher than linear powers of curvature) in the gravitational action is the inevitable consequence of models based on the spectral action principle.

The motivation for quantum field theory models of gravitational interactions with higher derivatives comes from many directions. Though there has been a recent revival of ideas related to higher derivative gravitational theories, the idea is not new. It was already observed by Utiyama and DeWitt \cite{Utiyama}, that the terms with higher derivatives of metric are needed for the renormalization of two-derivative matter theory on a curved background. (Basically, since in the matter part of the theory the couplings are with non-negative energy dimensions, the same energy dimensionality we must require from gravitational counter-terms. In $d=4$ spacetime dimensions, the generally covariant terms with dimensionless coupling constants must necessarily come with four derivatives of the metric.)  This is one of the reasons for higher derivative models. The renormalization of infinities in matter theory occurs in semi-classical approach when the gravity is taken to be a classical non-dynamical background and when the effects of back-reaction are neglected (so called quantum field theory in curved space).

On a different vain, also on the level of classical analysis, higher derivative theories come with some benefits \cite{stelle2}. When approximate solutions are studied (for example for Newtonian gravitational potential and using the Fourier transform method which presupposes some boundary conditions at spatial infinity) the effective classical spacetime shows no singularity at the origin (in this case there is no Newtonian-like $r^{-1}$ singularity of the gravitational potential \cite{Modesto:2014eta,Accioly:2016qeb,Giacchini:2016xns,Giacchini:2018gxp,Giacchini:2018wlf}). Similar resolution of singularities happens also  in cosmological framework \cite{Easson:2003ia,Biswas:2010zk}. However, these results have to be taken with a grain of salt, since they were obtained only in the perturbative scheme. Moreover, even in the linear approximation regime, the set of solutions of the higher derivative analogue of the Laplace equation still contains the singular $r^{-1}$ solution (as this was pointed out in \cite{exactsol}). Consequently, also the class of exact solutions is bigger (than in the standard two-derivative theory). And there typically among standard singular solutions (known from the two-derivative Einstein-Hilbert theory) we find also new types of solutions: often run-away or non-singular solutions. And these new solutions are responsible for resolution of singularity or give rise to cosmological inflation \cite{Starobinsky:1980te,Starobinsky:1981vz,Starobinsky:1983zz,cosmol}, where apparently they are not unwanted solutions anymore.  But we emphasize once again that in the whole set of exact solutions of higher derivative gravitational theories we have both singular solutions and new solutions, and the proper choice between them is specified by the boundary conditions. We just remark that in higher derivative theories we need to provide more boundary (or initial for dynamics in time, so for evolution) conditions than just in two-derivative theory, because of the increased number of derivatives on the metric field.

However, the main reason for higher derivatives in the gravitational setup has to do with quantum physics. As already noted by Stelle \cite{stelle1} the quantum gravity theory based on the four-derivative action in $d=4$ behaves much better regarding the situation with UV-divergences. Stelle's quadratic gravity is the first model of multiplicatively renormalizable gravitational theory. The quantization here can be performed using the standard methods in a fully covariant fashion (unlike in the more recent re-incarnation of the higher derivative gravity in the form of Ho\v{r}ava-Lifshitz gravity \cite{Horava:2009uw}, which explicitly breaks 4-dimensional diffeomorphisms,\footnote{For some applications of the methods of spectral geometry to Ho\v{r}ava-Lifshitz type theories, see \cite{Pinzul:2010ct,Mamiya:2013wqa,Pinzul:2014mva} and especially \cite{Lopes:2015bra,Pinzul:2016dwy} for the spectral action approach.} and where the higher derivatives are only in the spatial part). One sees that the inclusion of higher derivative terms as bare terms of the action is sufficient for absorption of divergences present in gravitational one-loop counter-terms \cite{hove}. This is intimately related to the observation made in quantum field theory in curved space, where we need the same counter-terms but there they are generated from the matter side \cite{Utiyama}. Using the Batalin-Vilkovisky formalism \cite{Batalin1,Batalin2} the proof of renormalizability can be extended to all perturbative loop orders. This means that all infinities appearing in the perturbative calculations can be consistently taken care of by re-defining the bare couplings of the theory which are present already at the tree-level. As another advantage one notices that such theory is predictive and only these couplings, which are present in the tree-level action, require experimental input, but going to higher loops does not force us to introduce new couplings at all. This is in strong distinction with quantum gravity theory based on E-H action, which is non-renormalizable from two loops on (at one-loop level without matter and on-shell this theory is miraculously finite), hence this theory is very badly non-predictive.

Moreover, this is not the end of good points about higher derivative models. In the quadratic gravity of Stelle the renormalization group (RG) behavior of the essential couplings of the theory can be studied. As it was first discussed by Fradkin and Tseytlin \cite{FT1,FT2} in $d=4$, the theory shows asymptotic freedom in all such couplings. This in turn signifies that the bare values for all couplings vanish (in the ultra-violet (UV) limit) and there is no issue of initial values for them, very similarly to the situation in QCD. Furthermore, the analogies of quadratic (in gravitational curvatures) gravity with quadratic (in YM field strengths)  matter \newpage \noindent theory \footnote{We remark that the standard Yang-Mills theory is a quadratic though \emph{two-derivative} renormalizable~theory.} go even further regarding, for example, the form of scattering amplitudes \cite{scattering}. Both theories are renormalizable and both are asymptotically-free. Finally, if one couples such matter theory to the quadratic gravity then also the total quantum system is described by renormalizable and asymptotically-free theory, where all the UV-infinities are under control. Such generalized framework provides a very interesting quantum laboratory for study of grand unified theories (GUT) coupled to gravity as field theory models of gravity-gauge unification (compare for example \cite{BOS}). In the connection with the spectral action approach, it is worth noting that from the point of view of (\ref{spectralaction}) both, YM and four-derivative gravity terms, appear exactly at the same level of the asymptotic expansion of the spectral action. Namely, they come from the Seeley-DeWitt coefficient $a_4$ (see the next section \ref{sec2}).

The four-derivative models can be also further generalized. Inclusion of terms with more derivatives is one such direction. When 6-derivative terms are added and enough care is exerted the gravitational model can still be shown to be renormalizable. But even more, for a particular form of generalization, the theory reveals to possess bigger control over perturbative divergences and it is said to be super-renormalizable. As proven by Asorey, Lopez and Shapiro \cite{shapiro3} for the case of a theory with six derivatives in $d=4$, all the diagrams with more than 3 loops are perfectly UV-finite. (The increase of number of derivatives can lower this bound and for example for theory with 10 derivatives in $d=4$ spacetime dimensions the UV-divergences remain only at the one-loop level \cite{superrenfin}.) The requirement of renormalizability constrains the possible terms, which could be added to the action of a six-derivative theory. First, in the action there must be terms quadratic in curvature in order to have a highly improved behavior of the flat spacetime propagator in the UV-regime. On the other hand, we can add also terms containing three curvatures but  for keeping the dimensionality of these terms under control they must be with no covariant derivatives in their generally covariant construction. This is in agreement with the scheme provided by effective field theories for gravity, where after inclusion of terms with four derivatives we, in principle, in our effective action should include all possible terms with six derivatives (partial, of the metric as seen in EOM). The possible terms can be quadratic or cubic in curvatures. We notice that such a theory naturally arises if we study the further expansion of the spectral action, namely the $a_6$ coefficient. However, such higher derivative gravitational theory based on the expansion of the spectral action up to $a_6$ coefficient has very particular coefficients in front of curvature invariants, which are dictated by spectral principle.  And therefore this special theory can be viewed as an example of a general six-derivative theory in $d=4$. In this sense we exploit here the predictive power of the spectral principle, since all these coefficients are unambiguously determined (up to the overall scaling, see in the next section \ref{sec2}). The main purpose of this paper is to study some (if any) consequences of this special form of the higher derivative (up to six-derivative) gravity.

The plan of the paper is as follows. In the next section \ref{sec2} we discuss the expansion of the purely geometric part of the spectral action (given by the first term in (\ref{spectralaction}) for the standard choice of the Dirac operator) up to six derivatives of the metric. Section \ref{EOM} contains our main results. It is devoted to the detailed study of the obtained higher derivative action in two different ``bases'': Weyl-- and Riemann--dominated. We also obtain the corresponding equations of motion. In section \ref{quantum_beta}, on the example of the beta function for the cosmological constant, we initiate the study of the consequences of the spectral action approach on the quantum level. We conclude with the extensive discussion of further possible consequences of our approach as well as delineation of further steps. Because one of the goals of the paper is to bring together the communities working in the areas of higher derivative gravity and non-commutative geometry, we include several appendices, where we provide the technical details of the construction (even though sometimes they are pretty standard) to make the paper as much self-contained as possible.

\hypersetup{bookmarksdepth=-2}
\pdfbookmark[1]{2 Spectral action}{name4}
\section{Spectral action}
\label{sec2}
\hypersetup{bookmarksdepth}

As we mentioned in the introduction, the full spectral action describes both, geometric and matter, sectors. Because we are interested in the case of pure gravity, here we will consider in details only the first term of (\ref{spectralaction}) for the standard choice of the Dirac operator. Before we proceed, the word of caution is in order. The whole construction is well-defined (or rather well-understood) only for the case of compact Riemannian manifolds. Hence, the final results for the Lorentzian signature should be understood as an analytic continuation of the results in the Euclidean framework (preceded by taking some decompactifying limit). Or, another point of view could be taken: because all the expressions are written in terms of the geometric invariants, one can continue formally to use them for the pseudo-Riemannian case (only carefully keeping track of various signs).

First of all, we have to fix our Dirac operator. The choice of the standard one could be motivated from two different points of view. The idea of the spectral action is that the same Dirac operator is used in both terms of (\ref{spectralaction}). This was used to some extreme in \cite{Lopes:2015bra,Pinzul:2016dwy} to relate the free parameters of the matter and gravitational sectors in Ho\v{r}ava-Lifshitz gravity. Because in the present paper, we do not want to modify the minimal coupling of matter to gravity and this is given by the Dirac-type second term in (\ref{spectralaction}), we must immediately conclude that we have to use the same, undeformed standard Dirac operator on the geometric side. On the other hand, there is a very powerful result in the spectral (a.k.a. non-commutative) approach to commutative geometry, due to Connes \cite{Connes:2008vs}, which, for our purposes, could be stated as follows: there is one to one correspondence between compact Riemannian geometries and the commutative spectral triples $(\mathcal{A},\mathrm{D},\mathcal{H})$ defined by the usual Dirac operator. All of these arguments lead us to the following choice (see appendix \ref{App:Lichnerowicz} for the notations and details of some relevant calculations)
\be
\mathrm{D} = \gamma^\mu (\partial_\mu - \omega_\mu)=: \gamma^\mu \nabla^\omega_\mu\ ,
\ee
where $\omega_\mu$ is the usual spin-connection.  Then the object we want to study is given by the first term in (\ref{spectralaction})
\be\label{spectralaction_grav}
\Tr\chi\left(\frac{\mathrm{D}^2}{\Lambda^2}\right).
\ee
It is obvious that the spectral action (\ref{spectralaction_grav}) is a highly non-trivial object, in particular it is non-local. So, is there any chance to have an explicit form of it? As we review in the appendix \ref{app_trace}, this could be done at least in some limit, using the so-called heat kernel expansion, which corresponds to the derivative expansion of (\ref{spectralaction_grav}). Using the general result (\ref{general_trace}), we can easily write the asymptotic expansion for (\ref{spectralaction_grav}) in $d=4$ Euclidean space dimensions
\be\label{heat_Dirac}
\Tr\chi\left(\frac{\mathrm{D}^2}{\Lambda^2}\right) = \sum_{k=0}^{\infty} \Lambda^{4-k} f_{2k} a_{2k}(\mathrm{D}^2)\ ,
\ee
where
\be\label{f}
f_0 = \!\int^\infty_0\! p \chi (p)\,dp\ ,\ f_2 = \!\int^\infty_0\! \chi (p)\,dp\,,\,\,\,  f_{2(2+k)}=(-1)^k \chi^{(k)}(0)\,\,\,
\,\,\,{\rm for}\,\,k\geqslant 0 \ .
\ee
Here $a_{2k}(\mathrm{D}^2)$ are Seeley-DeWitt coefficients for the elliptic operator
\be\label{Lichnerowicz}
-\mathrm{D}^2 = g^{\mu\nu}\nabla^\omega_\mu \nabla^\omega_\nu + \mathbb{E}\ ,
\ee
where $\mathbb{E}$ is some endomorphism of the corresponding bundle (see below and appendix \ref{App:Lichnerowicz}).

To determine these coefficients, the general method, due to Gilkey \cite{Gilkey:1995mj,Gilkey:2004dm}, or due to DeWitt \cite{DeWitt:1965} can be used. The essence of the former method could be roughly described as follows.

Because the operator (\ref{Lichnerowicz}) transforms covariantly under the diffeomorphisms of space and, in general, twisted spinor bundle endomorphisms, it is possible to show that all the coefficients in the heat kernel expansion (\ref{heat_kernel}) (or (\ref{heat_Dirac}) for the case of the interest) are given by the space volume integrals of the local geometric invariants, $a_n (x)$, and subsequent traces over the bundle indices (in the pure Riemannian case, this is a spinor bundle). Moreover, the whole explicit dependence on the dimension, $d$, of the manifold $\cal M$ is given by an overall factor, according to the formula:
\be
a_n (\mathrm{D}^2) = \frac{1}{(4\pi)^{d/2}}\!\int_\mathcal{M}\! \Tr a_n (x,\mathrm{D}^2)\sqrt{g}\, d^d x \ ,
\ee
which is essentially the formula (\ref{a_n}) from the appendix \ref{app_trace}. In a sense, the volume integrals of the expressions traced over internal indices can be viewed as the result of taking functional traces, both in internal and external space at one stroke. Therefore this generalizes the notion of trace to include also volume integrals and is in accord with the DeWitt convention \cite{DeWitt:1965} for compact index notation, treating on the same footing both spacetime $\cal M$ and internal space. The local invariants, $a_n(x)$, are constructed from the only geometric objects available: the Riemann curvature tensor $R_{\mu\nu\rho\sigma}$, the endomorphism $\mathbb{E}$ (\ref{Dirac_square}), the ``field strength'' of the bundle connection $\Omega_{\mu\nu}$ (\ref{Omega}) and the covariant derivatives $\nabla_\mu$. Here we use only Levi-Civita covariant derivatives $\nabla_\mu$, in opposition to total covariant derivative with spin-connection $\nabla_\mu^\omega$, because all geometric objects available are proportional to the identity $\mathds{1}$ in the spinor indices space (see also later and in the appendix \ref{App:Lichnerowicz}).
What invariants enter at each order $n$ of the expansion, $a_n (x,\mathrm{D}^2)$, can be easily decided from the dimensional analysis, after assigning the standard dimensions ($[x^\mu]=-1,\ [g_{\mu\nu}]=0$ and the rest follows) and requiring that the exponent of the heat kernel, $t \mathrm{D}^2$, in (\ref{hk0}) is dimensionless. From this point of view, it is clear that the heat kernel expansion (\ref{heat_kernel}) is a derivative expansion (rather than the curvature expansion - by integrating by parts, one can always trade some curvature for derivatives, see also below). In this way, each $a_n (x,\mathrm{D}^2)$ is a linear combination of the local geometric invariants, each having exactly $n$ derivatives, and the coefficients of this combination are universal, i.e. do not depend neither on geometry nor on the dimension $d$. This allows to fix these coefficients by evaluating the heat kernel on special geometries characterized by some degree of symmetry (like tori, spheres, etc.). This is the essence of the Gilkey method. However, the method by DeWitt is more algorithmic but more tedious since all these coefficients (generalized Schwinger coefficients) are obtained by differentiation of the world-line function and taking the coincidence limits \cite{DeWitt:1965, BVreport}. Just a word for terminology: we will call by $a_n (x,\mathrm{D}^2)$ the unintegrated coefficients, while the coefficients $a_n (\mathrm{D}^2)$ as in \eqref{heat_Dirac} are already integrated (and traced over). Using any of these methods, one can find the following expressions for the several first unintegrated Seeley-DeWitt coefficients \cite{gilkey,Gilkey:1995mj}:
\begin{allowdisplaybreaks}
\bea\label{Gilkey a}
a_{0}(x,\mathrm{D}^2)&=&{\Tr \mathds{1}} \label{Gilkey a0}\\
a_{2}(x,\mathrm{D}^2)&=&\frac{1}{6}{\Tr}\left\{ R + 6\mathbb{E} \right\} \label{Gilkey a2}\\
a_{4}(x,\mathrm{D}^2)&=&\frac{1}{360}{\Tr}\left\{ 12 \square R + 5R^2 - 2R_{\mu\nu}R^{\mu\nu} + 2R_{\mu\nu\rho\sigma}R^{\mu\nu\rho\sigma} + 60 R\mathbb{E} + \right.\nonumber \\
&& \left.+60\square \mathbb{E} + 180\mathbb{E}^2 + 30\Omega_{\mu\nu}\Omega^{\mu\nu} \right\} \label{Gilkey a4}\\
a_{6}(x,\mathrm{D}^2)&=&\frac{1}{360}{\Tr}\left\{ \frac{1}{14}\left(18\square^2 R+17R_{;\mu}R^{;\mu}-2R_{\mu\nu ;\rho}R^{\mu\nu ;\rho}-4 R_{\mu\nu ;\rho}R^{\mu\rho ;\nu}+28R\square R+\right.\right.\nn\\
&&\left. +9R_{\mu\nu\rho\sigma ;\kappa}R^{\mu\nu\rho\sigma ;\kappa}-8R_{\mu\nu}\square R^{\mu\nu}+24R_{\mu\nu}R^{\mu\rho}{}^{;\nu}{}_{;\rho}+12R_{\mu\nu\rho\sigma}\square R^{\mu\nu\rho\sigma}\right)+\nn\\
&&+\frac{1}{126}\left(35R^3-42RR_{\mu\nu}R^{\mu\nu}+42RR_{\mu\nu\rho\sigma}R^{\mu\nu\rho\sigma}-208R_{\mu\nu}R^{\mu}{}_{\rho}R^{\nu\rho}-\right.\nn\\ &&-192R_{\mu\nu}R_{\rho\sigma}R^{\mu\rho\nu\sigma}-48R_{\mu\nu}R^{\mu}{}_{\rho\sigma\kappa}R^{\nu\rho\sigma\kappa}-44R_{\mu\nu\rho\sigma}R^{\mu\nu}{}_{\kappa\lambda}R^{\rho\sigma\kappa\lambda}-\nn\\ &&\left.-80R_{\mu\nu\rho\sigma}R^{\mu}{}_{\kappa}{}^{\rho}{}_{\lambda}R^{\nu\kappa\sigma\lambda}\right)+ 8\Omega_{\mu\nu ;\rho}\Omega^{\mu\nu ;\rho}-2\Omega_{\mu\nu}{}^{;\mu}\Omega^{\nu\rho}{}_{;\rho}+12\Omega_{\mu\nu}\square \Omega^{\mu\nu}+\nn\\
&&+12\Omega_{\mu\nu}\Omega^{\mu}{}_{\rho}\Omega^{\nu\rho}-6R_{\mu\nu\rho\sigma}\Omega^{\mu\nu}\Omega^{\rho\sigma}-4R_{\mu\nu}\Omega^{\mu}{}_{\rho}\Omega^{\nu\rho}+
5R\Omega_{\mu\nu}\Omega^{\mu\nu}+\nn\\
&&+6\square^2 \mathbb{E}+60\mathbb{E}\square \mathbb{E}+30\mathbb{E}_{;\mu}\mathbb{E}^{;\mu}+60\mathbb{E}^{3}+30\mathbb{E}\Omega_{\mu\nu}\Omega^{\mu\nu}+10R\square \mathbb{E}+\nn\\
&& +4R_{\mu\nu}\mathbb{E}^{;\mu\nu}+12R_{;\mu}\mathbb{E}^{;\mu} + 30\mathbb{E}^{2}R+\nn\\
&& \left.+12\mathbb{E}\square R+5\mathbb{E}R^2-2\mathbb{E}R_{\mu\nu}R^{\mu\nu}+2\mathbb{E}R_{\mu\nu\rho\sigma}R^{\mu\nu\rho\sigma}\right\},\label{Gilkey a6}
\eea
\end{allowdisplaybreaks}
\hspace{-0.4cm} where $R$ is understood as $R \,\mathds{1}$ and we already used some trivial simplifications compared to \cite{Gilkey:1995mj,Gilkey:2004dm} due to the fact that the endomorphism $\mathbb{E}$ is proportional to the identity $\mathds{1}$ in the spinor space, that allowed to combine some terms, which otherwise are not equal. To further simplify these general formulas for $a_n (x,\mathrm{D}^2)$ we use the explicit expressions for $\mathbb{E}$ and $\Omega_{\mu\nu}$, from (\ref{Dirac_square}) and (\ref{Omega}),  as well as some standard trace identities for the Dirac gamma matrices (for our notations, see the appendix \ref{apA}):
\bea
&&\mathbb{E}=-\frac{1}{4}R\,\mathds{1}\ , \ \Omega_{\mu\nu}=\frac{1}{4}R_{\mu\nu}{}^{\rho\sigma}\gamma_{\rho\sigma}\ , \nonumber\\
&&\Tr \gamma_{\mu\nu} = 0\ , \ \Tr(\gamma_{\mu\nu} \gamma_{\rho\sigma}) = - 2\Tr(\mathds{1}) g_{\mu [\rho}g_{\sigma ]\nu}\ , \nonumber\\
&&\Tr(\gamma_{\mu\nu} \gamma_{\rho\sigma} \gamma_{\kappa\lambda}) = - 8\Tr(\mathds{1}) g_{[\![\rho [\mu}g_{\nu ][\kappa}g_{\lambda ]\sigma ]\!]} \ .
\eea
Using these relations, one can easily establish
\bea
&&\Tr(\Omega_{\mu\nu} \Omega_{\rho\sigma}) = - \frac{1}{8}\Tr(\mathds{1}) R_{\mu\nu}{}^{\kappa\lambda}R_{\rho\sigma\kappa\lambda} \ , \nonumber\\
&&\Tr(\Omega_{\mu\nu} \Omega_{\rho\sigma} \Omega_{\kappa\lambda}) = - \frac{1}{8}\Tr(\mathds{1}) R_{\mu\nu}{}^{\alpha\beta}R_{\rho\sigma\alpha\delta}R_{\kappa\lambda\beta}{}^{\delta} \ .
\eea
With the help of these relations, it is straightforward to re-write (\ref{Gilkey a0}-\ref{Gilkey a6}) in the following form
\begin{allowdisplaybreaks}
\bea
a_{0}(x,\mathrm{D}^2)&=&{\Tr \mathds{1}} \label{Gilkey a0R}\\
a_{2}(x,\mathrm{D}^2)&=&-\frac{1}{12}{\Tr}(\mathds{1})R \label{Gilkey a2R}\\
a_{4}(x,\mathrm{D}^2)&=&-\frac{1}{360}{\Tr}(\mathds{1})\left\{ 3 \square R - \frac{5}{4}R^2 + 2R_{\mu\nu}R^{\mu\nu} + \frac{7}{4}R_{\mu\nu\rho\sigma}R^{\mu\nu\rho\sigma} \right\}=\nonumber \\
&=& -\frac{1}{360}{\Tr}(\mathds{1})\left\{ 3 \square R - 3R^2 + 9R_{\mu\nu}R^{\mu\nu} + \frac{7}{4}{\rm GB}_0 \right\} \label{Gilkey a4R}\\
a_{6}(x,\mathrm{D}^2)&=&{\Tr}(\mathds{1})\left\{ -\frac{1}{1680}\square^2 R + \frac{1}{1440}R\square R + \frac{1}{4032}R_{;\mu}R^{;\mu} -\frac{1}{360}R^{\mu\nu}R_{;\mu\nu}-\right. \nonumber\\
&&-\frac{1}{560}R_{\mu\nu;\rho}R^{\mu\nu;\rho}+\frac{1}{1680}R_{\mu\nu;\rho}R^{\mu\rho;\nu}-\frac{1}{630}R_{\mu\nu}\square R^{\mu\nu}+\frac{1}{210}R_{\mu\nu}R^{\mu}{}_{\rho}{}^{;\nu\rho}- \nonumber \\
&&-\frac{1}{1008}R_{\mu\nu\rho\sigma;\kappa}R^{\mu\nu\rho\sigma;\kappa} - \frac{1}{560}R_{\mu\nu\rho\sigma}\square R^{\mu\nu\rho\sigma}
-\frac{1}{10368}R^{3} -\frac{13}{2835}R_{\mu\nu}R^{\mu}{}_{\rho}R^{\nu\rho}-\nn\\
&&-\frac{4}{945}R_{\mu\nu}R_{\rho\sigma}R^{\mu\rho\nu\sigma} +\frac{101}{90720}R_{\mu\nu\rho\sigma}R^{\mu\nu}{}_{\kappa\lambda}R^{\rho\sigma\kappa\lambda}+\frac{109}{45360}R_{\mu\nu\rho\sigma}R^{\mu}{}_{\kappa}{}^{\rho}{}_{\lambda}R^{\nu\kappa\sigma\lambda} +\nn\\
&&\left.+\frac{1}{3024}R_{\mu\nu}R^{\mu}{}_{\rho\kappa\lambda}R^{\nu\rho\kappa\lambda}+\frac{1}{2160}RR_{\mu\nu}R^{\mu\nu} +\frac{7}{17280}RR_{\mu\nu\rho\sigma}R^{\mu\nu\rho\sigma}\right\},\label{Gilkey a6R}
\eea
\end{allowdisplaybreaks}
\hspace{-0.2cm}where we have used the notation ${\rm GB}_0$ for the standard Gauss-Bonnet term (see the appendix \ref{apA} for more details). The above derivation is valid in general dimension $d$.

\pdfbookmark[1]{3 Higher derivative action and equations of motion}{name5}
\hypersetup{bookmarksdepth=-2}
\section{Higher derivative action and equations of motion}\label{EOM}
\hypersetup{bookmarksdepth}
\pdfbookmark[2]{3.1 Spectral action in different representations}{name6}
\hypersetup{bookmarksdepth=-2}
\subsection{Spectral action in different representations}
\hypersetup{bookmarksdepth}
\label{subsec31}

Combining (\ref{Gilkey a0R}-\ref{Gilkey a6R}) and the general result for the asymptotic expansion of the spectral action (\ref{heat_Dirac}), we arrive at the following action for the 6-derivative gravity:
\begin{eqnarray}\label{grav action 0}
S_{{\rm grav}} &=& \int\!d^{4}x\sqrt{g}\left[ \Lambda^4 \mu_0 - \Lambda^2 \mu_1 R + \mu_2 \left( R^2 - 3R_{\mu\nu}R^{\mu\nu} - \frac{7}{12}{\rm GB}_0 \right) \right. + \nonumber \\
&+& \frac{\mu_3}{\Lambda^2}\left\{ \frac{1}{2240}R\square R+\frac{1}{5040}R_{\mu\nu}\square R^{\mu\nu}-\frac{1}{1260}R_{\mu\nu\rho\sigma}\square R^{\mu\nu\rho\sigma}+\right.\nonumber\\
&+&\frac{1}{240}R_{\mu\nu}R^{\mu}{}_{\rho}{}^{;\nu\rho}-\frac{1}{360}R_{\mu\nu}R^{;\mu\nu}-\frac{1}{10368}R^{3}-\nonumber\\
&-&\frac{13}{2835}R_{\mu\nu}R^{\mu}{}_{\rho}R^{\nu\rho}-\frac{4}{945}R_{\mu\nu}R_{\rho\sigma}R^{\mu\rho\nu\sigma}+ \frac{101}{90720}R_{\mu\nu\rho\sigma}R^{\mu\nu}{}_{\kappa\lambda}R^{\rho\sigma\kappa\lambda}+\nonumber\\
&+&\frac{109}{45360}R_{\mu\nu\rho\sigma}R^{\mu}{}_{\kappa}{}^{\rho}{}_{\lambda}R^{\nu\kappa\sigma\lambda}+\frac{1}{3024}R_{\mu\nu}R^{\mu}{}_{\rho\sigma\kappa}R^{\nu\rho\sigma\kappa}-\nonumber\\
&-&\left.\left.\frac{1}{2160}RR_{\mu\nu}R^{\mu\nu}+\frac{7}{17280}RR_{\mu\nu\rho\sigma}R^{\mu\nu\rho\sigma}\right\}\right] \ ,
\end{eqnarray}
where we already integrated by parts and discarded all surface integrals. In fact, as we commented in the appendix \ref{app_trace} (see the footnote \ref{footnote}), we could do it because the expansion (\ref{heat_Dirac}) is already written under the assumption that there is no boundary. Also we explicitly kept the dependence on the cut-off scale $\Lambda$, while calling the numerical coefficients for each level with $2k$ derivatives by $\mu_k$ (which anyway are not fixed by the model). In this regard, three comments are in order.

1) It is clear either from (\ref{heat_Dirac}) or from (\ref{grav action 0}) that the level with $2k$ derivatives is suppressed by the factor of $\Lambda^{-2}$ compared to the one with $2(k-1)$ derivatives. This seems very natural except that the term without derivatives should correspond to the cosmological constant term, which would lead to the huge dark energy density. We will not address this point in this work and rather refer to \cite{Chamseddine:2008zj} where the way to resolve this problem is discussed.

2) As we said above, the numerical coefficients $\mu_k$ are not specified within this approach and, in principle, are free parameters of the model. Still one ``prediction'' can be made. Let us recall that the arbitrary function $\chi(p)$ in (\ref{spectralaction_grav}) is supposed to be some kind of a cut-off function and the coefficients $\mu_k$ are proportional to $f_{2k}$ given by (\ref{f}) in terms of this cut-off function. From this it is -clear that while $\mu_k$, $k=0,1,2$ are really arbitrary non-zero numbers, $\mu_3$, being proportional to $\chi'(0)$, should be zero if $\chi$ is flat at the beginning of the spectrum (as it is in the case of the standard cut-off functions). As we commented above, the actual shape of $\chi$ should be fixed by some fundamental theory, but we might expect that the behavior of $\chi$ in IR is very close to the flat one, i.e. that $\chi'(0)\ll 1$. This is due to the fact that the spectrum of the standard Dirac operator controls the classical geometry of spacetime and we do not want to distort this spectrum too much (by modulating it with $\chi$) in IR. Hence, based on this discussion, the prediction of the model would be the additional suppression of the 6-derivative term by the small factor $\mu_3 \ll 1$ (in addition to $\Lambda^{-2}$ suppression).

3) The real prediction of this approach is given by the values of the coefficients \textit{within} each derivative level (so the relative weights of terms). This drastically reduces the number of free parameters in higher derivative gravity. E.g., without spectral action, the number of free parameters for higher derivative gravity with up to six derivatives would be well above 10, while in our model we have just 4 (these are $\mu_0$, $\mu_1$, $\mu_2$, and $\mu_3$ respectively).

The main goal of this section is to maximally simplify the action (\ref{grav action 0}) and present it in several equivalent forms that might be useful for different types of problems.

As the first step, let us make the most obvious simplifications related to the following terms $R_{\mu\nu}R^{\mu}{}_{\rho}{}^{;\nu\rho}$ and $R_{\mu\nu}R^{;\mu\nu}$ in (\ref{grav action 0}). One trivially has
\begin{eqnarray}
R_{\mu\nu}R^{\mu}{}_{\rho}{}^{;\nu\rho}=2 R_{\mu\nu}R^{\mu}{}_{\rho}{}^{;[\nu\rho]} + R_{\mu\nu}R^{\mu}{}_{\rho}{}^{;\rho\nu} =- R_{\mu\nu}R^{\rho\sigma}R_{\rho}{}^{\nu\mu}{}_{\sigma} + R_{\mu\nu}R^{\nu\rho}R^{\mu}{}_{\rho} + R_{\mu\nu}R^{\mu}{}_{\rho}{}^{;\rho\nu} \ , \nonumber
\end{eqnarray}
where we used the standard result for the commutator of the covariant derivatives (with the sign conventions from appendix \ref{apA}). Now using the doubly contracted second Bianchi identity (\ref{Contrct. Sec. Bianchi 2_Ric}), $R^{\mu}{}_{\nu ;\mu}= \frac{1}{2}R_{;\nu}$ we have for the integrals (again, discarding total derivatives):
\begin{eqnarray}
\int d^{d}x\sqrt{g}R_{\mu\nu}R^{\mu}{}_{\rho}{}^{;\nu\rho}&=&\int d^{d}x\sqrt{g}\left(\frac{1}{4}R\square R + R_{\mu\nu}R_{\rho\sigma}R^{\mu\rho\nu\sigma} + R_{\mu\nu}R^{\mu}{}_{\rho}R^{\nu\rho}\right),\label{RR1}\\
\int d^{d}x\sqrt{g}R_{\mu\nu}R^{;\mu\nu}&=&\frac{1}{2}\int d^{d}x\sqrt{g}R\square R\ . \label{RR2}
\end{eqnarray}
The equation (\ref{RR1}) is a typical example of how we can trade derivatives for curvature, so we again stress that the heat kernel expansion is a derivative expansion rather than the curvature one.

Now we would like to get rid of the term $R_{\mu\nu\rho\sigma}\square R^{\mu\nu\rho\sigma}$. Note that this term enters the generalized Gauss-Bonet term, ${\rm GB}_1$ (\ref{GB}). Repeatedly using commutators of the covariant derivatives and the contracted second Bianchi identity, as in the derivation of (\ref{RR1}) and (\ref{RR2}), one can easily get
\begin{eqnarray}\label{GB1_Int}
\int d^{d}x\sqrt{g}\,{\rm GB}_{1} &=& \int d^{d}x\sqrt{g}\left(-4R_{\mu\nu}R_{\rho\sigma}R^{\mu\rho\nu\sigma}-4R_{\mu\nu}R^{\mu}{}_{\rho}R^{\nu\rho} + 4R_{\mu\nu\rho\sigma}R^{\mu}{}_{\kappa}{}^{\rho}{}_{\lambda}R^{\nu\kappa\sigma\lambda}+\right.\nonumber\\ &+& \left. R_{\mu\nu\rho\sigma}R^{\mu\nu}{}_{\kappa\lambda}R^{\rho\sigma\kappa\lambda}+2R_{\mu\nu}R^{\mu}{}_{\rho\sigma\kappa}R^{\nu\rho\sigma\kappa}\right).
\end{eqnarray}

So far for our simplification of the action functional we used identities valid in any number of dimensions $d$. Final steps are done with the assumption $d=4$. Then further simplification is possible due to a very simple observation: in 4 dimensions, anti-symmetrizing any tensor with respect to five or more indices identically gives zero. Choosing different products of three Riemann, Ricci or Weyl tensors, this leads to the following identities \cite{vanNieuwenhuizen:1976vb,Harvey:1995}
\bea
&& R_{\mu\nu\rho\sigma}R^{\mu\nu}{}_{\kappa\lambda}R^{\rho\sigma\kappa\lambda}-2R_{\mu\nu\rho\sigma}R^{\mu}{}_{\kappa}{}^{\rho}{}_{\lambda}R^{\nu\kappa\sigma\lambda}+5R_{\mu\nu}R^{\mu}{}_{\rho\sigma\kappa}R^{\nu\rho\sigma\kappa}+ \nonumber\\
&& +4R_{\mu\nu}R_{\rho\sigma}R^{\mu\rho\nu\sigma}-2R_{\mu\nu}R^{\mu}{}_{\rho}R^{\nu\rho}-\frac{1}{2}RR_{\mu\nu\rho\sigma}R^{\mu\nu\rho\sigma}+RR_{\mu\nu}R^{\mu\nu}=0 \ ,\nonumber\\
&& 2R_{\mu\nu}R^{\mu}{}_{\rho\sigma\kappa}R^{\nu\rho\sigma\kappa}-\frac{1}{2}RR_{\mu\nu\rho\sigma}R^{\mu\nu\rho\sigma}-4R_{\mu\nu}R^{\mu}{}_{\rho}R^{\nu\rho}+4RR_{\mu\nu}R^{\mu\nu} +4R_{\mu\nu}R_{\rho\sigma}R^{\mu\rho\nu\sigma}-\frac{1}{2}R^{3}=0 \ ,\nonumber
\eea
or equivalently in terms of the traceless Weyl tensor
\bea\label{Identities_Harvey}
&& 4R_{\mu\nu}C^{\mu}{}_{\rho\sigma\kappa}C^{\nu\rho\sigma\kappa}-RC_{\mu\nu\rho\sigma}C^{\mu\nu\rho\sigma}=0 \ ,\nonumber\\
&& C_{\mu\nu\rho\sigma}C^{\mu\nu}{}_{\kappa\lambda}C^{\rho\sigma\kappa\lambda}-2C_{\mu\nu\rho\sigma}C^{\mu}{}_{\kappa}{}^{\rho}{}_{\lambda}C^{\nu\kappa\sigma\lambda}=0 \ .
\eea

Combining these relations with (\ref{RR1}), (\ref{RR2}) and (\ref{GB1_Int}) one gets, after some straightforward calculations, the following compact result for the action (\ref{grav action 0}):
\begin{eqnarray}\label{grav action R}
S_{{\rm grav}} &=& \int\!d^{4}x\sqrt{g}\left[ \Lambda^4 \mu_0 - \Lambda^2 \mu_1 R + \mu_2 \left( R^2 - 3R_{\mu\nu}R^{\mu\nu} -\frac{7}{12} {\rm GB}_0 \right) \right. + \nonumber \\
&& + \frac{\mu_3}{\Lambda^2}\left\{
\frac{9}{10}R\square R-3R_{\mu\nu}\square R^{\mu\nu}+8R_{\mu\nu}R_{\rho\sigma}R^{\mu\rho\nu\sigma}-\right.\nn\\&&-\frac{43}{15}R_{\mu\nu}R^{\mu}{}_{\rho}R^{\nu\rho}- \frac{9}{10}R^{3}+\frac{13}{2}RR_{\mu\nu}R^{\mu\nu}-\frac{1}{5}RR_{\mu\nu\rho\sigma}R^{\mu\nu\rho\sigma}-\nonumber\\ &&\left.\left.-\frac{1}{15}R_{\mu\nu\rho\sigma}R^{\mu\nu}{}_{\kappa\lambda}R^{\rho\sigma\kappa\lambda}\right\}\right] \ ,
\end{eqnarray}
where compared to (\ref{grav action 0}) we changed $\mu_3 \rightarrow 1008\, \mu_3 $ (though, as we mentioned, this is quite irrelevant taking into account that $\mu_3$ is a free parameter), also we kept the topological term ${\rm GB_0}$ even though it will not contribute to the classical equations of motion (and, of course, it should be kept for quantum calculations). This is exactly the form that we call the action in the \textit{Riemann basis} or the \textit{Riemann--dominated} action.

It is clear that the Riemann--dominated form is not the most convenient one if one wants to study the conformal backgrounds. This motivates us to look for the equivalent expression, but now written in the \textit{Weyl basis} or in the \textit{Weyl--dominated} form. This is readily done by expressing most of the terms in the Riemann--dominated action (\ref{grav action R}) with the help of the definition of the Weyl tensor (\ref{Weyl}) and the relation (\ref{Weyl box Weyl}) and, when necessary, again using (\ref{GB1_Int}) and the identities (\ref{Identities_Harvey}). After not so lengthy and straightforward manipulations we arrive at the result:
\begin{eqnarray}\label{grav action W}
S_{{\rm grav}} &=& \int\!d^{4}x\sqrt{g}\left[ \Lambda^4 \mu_0 - \Lambda^2 \mu_1 R - \frac{3\mu_2}{2}\left( C_{\mu\nu\rho\sigma}C^{\mu\nu\rho\sigma} -\frac{11}{18}{\rm GB}_0 \right)\right. + \nonumber \\
&& + \frac{\mu_3}{\Lambda^2}\left\{
-\frac{1}{10}R\square R-\frac{3}{2}C_{\mu\nu\rho\sigma}\square C^{\mu\nu\rho\sigma}-\frac{2}{135}R^{3}+\right.\nn\\ &&+\frac{1}{3}RR_{\mu\nu}R^{\mu\nu}-\frac{13}{15}R_{\mu\nu}R^{\mu}{}_{\rho}R^{\nu\rho}+\frac{7}{12}R C_{\mu\nu\rho\sigma}C^{\mu\nu\rho\sigma}+\nn\\ &&\left.\left. +\frac{23}{5}R_{\mu\nu}R_{\rho\sigma}C^{\mu\rho\nu\sigma}+\frac{133}{30}C_{\mu\nu\rho\sigma}C^{\mu\nu}{}_{\kappa\lambda}C^{\rho\sigma\kappa\lambda}\right\}\right].
\end{eqnarray}

Yet another form might be useful for comparing with other works on higher derivative gravity (see, e.g. \cite{shapiro3,RGsuperren}) where the action is written in ($R,C,{\rm GB}$)-basis:
\begin{eqnarray}\label{RCGB}
S_{HD} = S_{EH + \Lambda}+ \int\!d^{4}x\sqrt{g}\sum\limits_{k=0}^{N}\left( c^R_k R\square^k R + c^C_k C_{\mu\nu\rho\sigma}\square^k C^{\mu\nu\rho\sigma} + c^{\rm GB}_k {\rm GB}_{k} \right) + V({\cal R}) \ ,
\end{eqnarray}
where $V({\cal R})$ is some ``potential'' depending on the curvature $\cal R$.\footnote{\label{R_terms}As a generalized curvature $\cal R$ here we understand any tensor constructed from Riemann tensor $R_{\mu\nu\rho\sigma}$ by various contractions.} The case $V({\cal R}) = 0$ would correspond to the ``minimal'' action in this basis (but compare with the discussion in \cite{superrenfin,finmss}). But one should remember that in view of (\ref{GB1_Int}) it is \textit{not} preferred in any other sense. In any case, the spectral action in this basis takes the ``non-minimal'' form  with
\begin{eqnarray}\label{grav action GB}
S_{{\rm grav}} &=& \int\!d^{4}x\sqrt{g}\left[ \Lambda^4 \mu_0 - \Lambda^2 \mu_1 R - \frac{3\mu_2}{2}\left( C_{\mu\nu\rho\sigma}C^{\mu\nu\rho\sigma} -\frac{11}{18}{\rm GB}_0 \right)\right. + \nonumber \\
&& + \frac{\mu_3}{\Lambda^2}\left\{-\frac{1}{10}R\square R-\frac{3}{2}C_{\mu\nu\rho\sigma}\square C^{\mu\nu\rho\sigma}+\frac{41}{60}\mathrm{GB_{1}}+\right.\nn\\
&&+\frac{7}{60}R^{3}-\frac{19}{15}R_{\mu\nu}R^{\mu}{}_{\rho}R^{\nu\rho}+ \frac{47}{15}R_{\mu\nu}R_{\rho\sigma}R^{\mu\rho\nu\sigma}+\frac{86}{15}R_{\mu\nu}R^{\mu}{}_{\rho\sigma\kappa}R^{\nu\rho\sigma\kappa}+\nn\\ &&\left.\left.+\frac{143}{60}R_{\mu\nu\rho\sigma}R^{\mu\nu}{}_{\kappa\lambda}R^{\rho\sigma\kappa\lambda}\right\}\right].
\end{eqnarray}

After obtaining the simplified forms of the 6-derivative gravity coming from the spectral action, it is reasonable to ask whether we have any further advantages of the approach beyond the obvious rigidness of the result (in the sense of the great reduction in the number of free parameters). In other words: is there anything special about the spectral action approach from the point of view of higher derivative gravities? In this section we will touch on this point on the classical level and in the next we briefly discuss some quantum aspects postponing a more detailed discussion of the quantum case for the future work.

One may guess that the spectral action has a lot to do with additional symmetries present in the gravitational interactions. The conformal symmetry may play such a role. Actually, based on the explicit results of the $a_4$  coefficient (last part of the first line in \eqref{grav action W}), one would be almost convinced about this since in $d=4$ dimensions the only non-trivial term appearing there is the $C^2$ term, which transforms in a covariant way under local conformal transformations. (We neglect here the Gauss-Bonnet term ${\rm GB}_0$ since this is a topological term in $d=4$.) Only in four dimensions, in $a_4$ we have only $C^2$ and ${\rm GB}_0$ terms, in other dimensions there is a non-zero coefficient in
front of the $R^2$ term. Moreover, in $a_2$ we have only a term with Ricci scalar $R$ and this is
exceptionally conformally covariant term in the action of gravity in dimensions $d=2$. This hope is reinforced by the fact that the $\sqrt{g}R^2$ term is missing in $a_4$ in $d=4$ and this term is only globally scale-invariant in four-dimensional case (invariant only under rigid scale transformations) and hence dimensionless. In $a_6$ we naturally have terms with six derivatives, so they cannot be dimensionless in $d=4$, but they might transform covariantly (that is with a weight factor) under conformal transformations. The condition for this is that they would have to be built out of only Weyl tensor and its various contractions and no covariant derivatives or covariant box operators acting on these conformal tensors \cite{confreview}. Then they would be truly conformally invariant in $d=6$ dimensions and there they would be therefore dimensionless. However, the inspection of the action written in the Weyl--dominated form \eqref{grav action W} shows that this hope for additional symmetry of the spectral action is not fulfiled. We find there, in the sector of terms with six derivatives, terms built also with Ricci scalar (which does not transform conformally in a neat way) and even terms of the type $C_{\mu\nu\rho\sigma}\square C^{\mu\nu\rho\sigma}$, which would break conformal symmetry in $d=6$. Based on the explicit example of the $a_6$ coefficient we conclude that generally  conformal symmetry (even in a restricted sense in $d$ spacetime dimensions for the $a_d$ coefficient of the expansion) is not a feature of the spectral action approach.

As the first application of the simplified action in the Weyl basis, let us evaluate the action (\ref{grav action W}) on the conformal background, i.e. when $C_{\mu\nu\rho\sigma} = 0$ \footnote{By conformal backgrounds we mean backgrounds which are conformally flat, that is by conformal transformation of the metric tensor $g_{\mu\nu}\to g'_{\mu\nu}=\Omega^2(x)g_{\mu\nu}$ with some suitable function $\Omega(x)$ we get the metric $g'_{\mu\nu}$ as the metric of flat spacetime, i.e. the Riemann tensor of the $g'$ metric vanishes identically. The condition for conformal flatness in dimensions $d\geqslant4$ is equivalent to vanishing of Weyl tensor $C_{\mu\nu\rho\sigma}$. Hence this last tensor is also called as the tensor of conformal curvature.}. We want to compare this with the discussion in \cite{Chamseddine:2008zj} where this was done for the special case of $S^1 \times S^3$ background. Even for this case, the calculation was extremely complicated technically and the main result (that the 6-derivative part of the action is zero for this background, see below) was very surprising. Here we re-derive this result, which will also provide an independent check of our action (\ref{grav action W}), and discuss what can be said in the case of a general conformal background. In this way we generalize the results from \cite{Chamseddine:2008zj} including the impact of terms with covariant derivatives on curvature tensors. In the remainder of this subsection we analyze this issue, while in the next subsection we analyze whether some commonly known background spacetimes are exact solutions of the theory.

Let us trivially evaluate (\ref{grav action W}) for the geometries with $C_{\mu\nu\rho\sigma} = 0$.\footnote{Note that for the calculation of the equations of motion on the conformal background, one cannot just set all the Weyl terms to zero. This is because the variation of the Weyl tensor evaluated on the conformal background is not zero, so one has to keep the terms linear in $C_{\mu\nu\!\rho\sigma}$\,, see below.} The result is
\begin{eqnarray}\label{grav action W conformal}
S_{{\rm grav}}|_{\rm conf} &=& \int\!d^{4}x\sqrt{g}\left[ \Lambda^4 \mu_0 - \Lambda^2 \mu_1 R + \frac{11\mu_2}{12}{\rm GB}_0 \right. + \nonumber \\
&& + \frac{\mu_3}{\Lambda^2}\left\{
-\frac{1}{10}R\square R-\frac{2}{135}R^{3}
\left. +\frac{1}{3}RR_{\mu\nu}R^{\mu\nu}-\frac{13}{15}R_{\mu\nu}R^{\mu}{}_{\rho}R^{\nu\rho}\right\}\right].
\end{eqnarray}
Already from this result it is obvious that in the case of a general conformal background, i.e. when $C_{\mu\nu\rho\sigma} = 0$, the 6-derivative part of the action will not be zero. This makes the result for $S^1 \times S^3$ even more surprising. Our general result (\ref{grav action W conformal}) allows to obtain it almost trivially compared to \cite{Chamseddine:2008zj}. First of all, because this background has a constant scalar curvature (see (\ref{Ricci S1S3}) below), the term $R\square R$ drops out automatically. The only non-trivial components of Riemann tensor are
\begin{eqnarray}\label{Riemann S1S3}
R_{ijkl}= - \frac{1}{a^2}\left( g_{ik}g_{jl} - g_{il}g_{jk} \right),
\end{eqnarray}
where $a$ is the radius of $S^3$ and the space-like indices $i,j,k,l = 1,2,3$. Contracting, we get Ricci tensor and the scalar curvature (pay attention to our sign convention in \eqref{sriem}, \eqref{sric} and \eqref{signrs})
\begin{eqnarray}\label{Ricci S1S3}
R_{ij}&=& \frac{2}{a^2}g_{ij}\ \mbox{and the rest are zero},\nn \\
R &=& \frac{6}{a^2}\ .
\end{eqnarray}
Using this, one easily calculates the relevant terms in (\ref{grav action W conformal}).
\begin{eqnarray}\label{Terms S1S3}
&& R_{\mu\nu\rho\sigma}R^{\mu\nu\rho\sigma} = R_{\mu\nu}R^{\mu\nu}=\frac{12}{a^4}\ ,\ R^2 = \frac{36}{a^4}\ , \nn \\
&& {\rm GB}_0 =R_{\mu\nu\rho\sigma}R^{\mu\nu\rho\sigma}-4R_{\mu\nu}R^{\mu\nu}+R^{2}=0  \ ,\nn \\
&& \frac{13}{15}R_{\mu\nu}R^{\mu}{}_{\rho}R^{\nu\rho} = \frac{24}{a^6}\ .
\end{eqnarray}
Combining these results and using them in (\ref{grav action W conformal}), it is trivial to see that the 6-derivative term is zero, while the whole action evaluated on this background is given by
\begin{eqnarray}\label{grav action S1S3}
S_{{\rm grav}}|_{S^1\times S^3} &=&  4\pi^3 a^3 b \left( \Lambda^4 \mu_0 -\Lambda^2 \mu_1 \frac{6}{a^2} \right) \ ,
\end{eqnarray}
where $4\pi^3 a^3 b$ is just the volume of $S^1\times S^3$ with $b$ being the radius of $S^1$ and $a$ of $S^3$. The equation (\ref{grav action S1S3}) is essentially the result obtained in \cite{Chamseddine:2008zj} by the direct evaluation of $a_6$ (\ref{Gilkey a6}) for the $S^1\times S^3$ background. Thus our approach correctly reproduces this result and demonstrates the role of the performed simplifications. Also, we want to stress one more time that the cancellation of the 6-order terms for this background should be considered as accidental: it is not automatic but happens due to the non-trivial cancellation between terms depending on the curvature. Because this happens exactly for the coefficients fixed by the spectral action, one might speculate that the spectral action somehow prefers this background.

\hypersetup{bookmarksdepth=1}
\pdfbookmark[2]{3.2 Equations of motion}{name7}
\hypersetup{bookmarksdepth=-2}
\subsection{Equations of motion}
\label{subsec32}
\hypersetup{bookmarksdepth}

Now let us make one step further and derive the equations of motion for the Riemann-- and Weyl--dominated forms of the action, (\ref{grav action R}), (\ref{grav action W}). Though the general equations of motion following from (\ref{grav action R}) are not very illuminating, in the appendix \ref{AppEOM} we give the final result for them for the possible future references and applications. Below we will consider a special case of these equations for the very important class of the Ricci--flat backgrounds. In the case of standard GR, Ricci--flat geometries are the special types of the Einstein spaces for the case of zero cosmological constant. The most known (and, probably, the most important) of these solutions is the Schwarzschild one. In the framework of the higher derivative gravity, one would like to find the corrections to this solution (and extract from there the quantum-gravitational corrections to Newton's law). Postponing this very important task for the future research, here we just show how the simplified form of the action (\ref{grav action R}) easily allows to derive a compact set of the equations of motion for a general Ricci--flat background. Also we verify that, not surprisingly, the Schwarzschild metric is \textit{not} a vacuum solution of these equations and rather requires as a source the energy-momentum tensor with very peculiar, exotic and unphysical properties.

To derive the vacuum EOM (where we do not include matter energy-momentum tensor on the RHS) for $R_{\mu\nu} = 0$ case from (\ref{grav action R}) we note that one can set to zero in (\ref{grav action R}) all the terms that are more than linear in $R_{\mu\nu}$ and $R$ (but not in $R_{\mu\nu\rho\sigma}$\,!). The linear terms should be kept. One should also drop the ${\rm GB}_0$ term. This immediately kills almost all the terms in (\ref{grav action R}):
\begin{eqnarray}\label{grav action R linear}
S_{{\rm grav}} &=& \int\!d^{4}x\sqrt{g}\left[ \Lambda^4 \mu_0 - \Lambda^2 \mu_1 R  - \frac{\mu_3}{5\Lambda^2}\left(
RR_{\mu\nu\rho\sigma}R^{\mu\nu\rho\sigma}+\frac{1}{3}R_{\mu\nu\rho\sigma}R^{\mu\nu}{}_{\kappa\lambda}R^{\rho\sigma\kappa\lambda}\right)\right] + \nonumber \\
&& + \mathcal{O}\left( R^2 , R_{\mu\nu}^2, R R_{\mu\nu} \right).
\end{eqnarray}
Now (\ref{grav action R linear}) can be straightforwardly varied using the standard variations collected in the appendix \ref{AppEOM} (\ref{varR}) producing a very compact result for the tensor of equations of motion (sometimes called a bit incorrectly by generalized Einstein tensor) $E^{\alpha\beta} = \frac{1}{\sqrt{g}}\frac{\delta S_{\rm grav}}{\delta g_{\alpha\beta}}$. The tensor $E^{\alpha\beta}$ is the gravitational part of EOM of the system and it reads
\begin{eqnarray}\label{Ricci flat EOM}
E^{\alpha\beta}&=&\frac{\Lambda^4 \mu_0}{2}g^{\alpha\beta}
-\frac{\mu_3}{5\Lambda^2}\left[ \frac{1}{6}g^{\alpha\beta}R^{\mu\nu\rho\sigma}R_{\mu\nu}{}^{\kappa\lambda}R_{\rho\sigma\kappa\lambda}+4R^{\alpha\mu\nu\rho}R^{\beta\sigma}{}_{\nu}{}^{\kappa}R_{\mu\sigma\rho\kappa}+ \right.\nn \\
&&\left. +2R^{\alpha\mu\nu\rho;\sigma}R^{\beta}{}_{\sigma\nu\rho;\mu}-\nabla^\kappa\nabla^\lambda\left(\left[g^{\alpha\beta} g_{\kappa\lambda}-\delta^\beta{}_\kappa\delta^\alpha{}_\lambda \right]R^{\mu\nu\rho\sigma}R_{\mu\nu\rho\sigma} \right)\right]\ .
\end{eqnarray}
As we said above, not surprisingly, the standard Schwarzschild spacetime with a metric tensor in standard Schwarzschild coordinate system given by
\begin{eqnarray}\label{Schwarz}
ds^2 = - \left( 1 - \frac{2M}{r} \right)dt^2 + \left( 1 - \frac{2M}{r} \right)^{-1} dr^2 + r^2 d\Omega^2 \ ,
\end{eqnarray}
(by $d\Omega^2$, as usual, we denote angular part of the metric, that is $d\Omega^2=d\theta^2+\sin^2\theta d\phi^2$) is \emph{not} a vacuum solution to the equations (\ref{Ricci flat EOM}). We find for the respective components
\begin{eqnarray}\label{Schwarz EOM}
E^{t}{}_{t} &=& \frac{8\mu_3}{5\Lambda^2}\frac{M^{2}(-298M+135r)}{r^{9}}\ , \nn\\
E^{r}{}_{r} &=& \frac{56\mu_3}{5\Lambda^2}\frac{M^{2}(14M-9r)}{r^{9}}\ , \nn\\
E^{\theta}{}_{\theta} &=&E^{\phi}{}_{\phi}=\frac{8\mu_3}{5\Lambda^2}\frac{M^{2}(-442M+189r)}{r^{9}} \ .
\end{eqnarray}
\begin{figure} 
    \centering
            \includegraphics[scale=1.2]{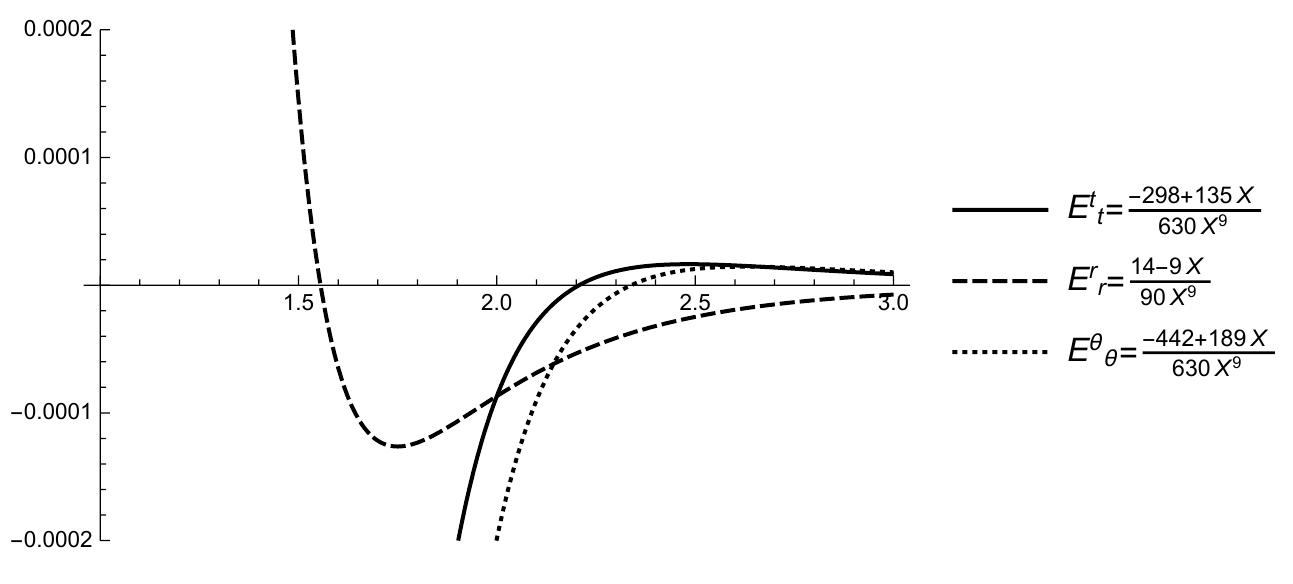}
    \caption{Various components of the energy-momentum tensor of matter source (or effectively of the LHS of the EOM) for a Schwarzschild ansatz as functions of the dimensionless radial coordinate $X$. For preparation of the plots we used an identification $\frac{8\mu_3M^3}{5\Lambda^2}\to\frac{1}{630}$.}\label{Schwz._plots}

\end{figure}
\hspace{-0.3cm} Utilizing (\ref{Schwarz EOM}) one can come up with the corresponding plots as shown in Fig. \ref{Schwz._plots}. There we exploited the dimensionless variable $X=\frac{r}{M}$ and rescaled the components of EOM in (\ref{Schwarz EOM}) by a common power $M^3$. It is evident that zeros for the three components $E^{t}{}_{t}$, $E^{r}{}_{r}$ and $E^{\theta}{}_{\theta}=E^{\phi}{}_{\phi}$ are given by approximate values of the $X$ coordinate $X\approx2.2,\,1.55\,\,\mathrm{and}\,\,2.34$, respectively. 
We observe that energy density $E^{t}{}_{t}$ and azimuthal pressure $E^{\theta}{}_{\theta}=E^{\phi}{}_{\phi}$ both are positive for considerably higher radii. Whereas, the radial pressure $E^{r}{}_{r}$ becomes negative at $X\gtrapprox1.55$. So we see that to get (\ref{Schwarz}) as a solution, (\ref{Ricci flat EOM}) must be sourced by a very non-physical energy momentum tensor (the cosmological constant term with $\mu_0$ in (\ref{Ricci flat EOM}) is set to zero).

One can check that the components of the effective energy-momentum tensor (as evaluated in Eqs. (\ref{Schwarz EOM})) do not satisfy energy conditions (neither strong, dominant, nor null one). This feature is actually common to almost all higher derivative theories since this is the price to have among bigger set of solutions also those which are non-singular. (This is a caveat to the celebrated Hawking-Penrose theorems about inevitability of spacetime singularities -- in theories with higher derivatives classical energy conditions are violated and that is why singularities can be avoided in some exact solutions of such theories.)

However, from the point of view of effective theory two aspects are worrisome here. First is that for $X\gtrapprox1.55$ the radial pressure $E^r{}_r$ attains negative values. This characteristics cannot be accepted as pertaining to an effective matter source since there does not exist any type of classical matter which exhibits negative pressure. Some exotic examples are brought by quantum effects (vacuum polarization effects) or vacuum energy realized for example as a cosmological constant source (or by Casimir effects). The second problem is not so severe since it touches on the behavior for smaller radii and for two different components of the effective energy-momentum tensor, namely for $E^t{}_t$ and $E^\theta{}_\theta=E^\phi{}_\phi$. We find that these components become negative inside the core of our solution. Hence our solution cannot be a physical representation of a star in higher derivative gravitational theories. This latter issue is not so problematic because these effects happen roughly under the classical Schwarzschild horizon, which is located at $X=2$. Even in Einstein--Hilbert gravitational theory the source of the gravitationally collapsing configuration (producing eventually a black hole) inside the Schwarzschild horizon is not a stationary matter source and energy densities there may be negatively valued.

Of course, what one should do, instead of just checking that (\ref{Schwarz}) is not a vacuum solution of this version of the higher derivative gravity, is to look for the corrections to the Schwarzschild metric following from the full set of the equations (\ref{EOM general}) (now one cannot use the Ricci--flat ansatz). But this is technically quite involved and will require some numerical study. We are planning on returning to this in the future research.

Analogous analysis can be performed for the Weyl--dominated form of the action (\ref{grav action W}) in the case of a conformal background. Again, in this case one can keep in (\ref{grav action W}) only the terms up to the first order in $C_{\mu\nu\rho\sigma}$ (and one can drop ${\rm GB}_0$ as it will not contribute to the equations of motion):
\begin{eqnarray}\label{grav action W linear}
S_{{\rm grav}} &=& \int\!d^{4}x\sqrt{g}\left[ \Lambda^4 \mu_0 + \right. \frac{\mu_3}{5\Lambda^2}\left\{ -\frac{1}{2}R\square R-\frac{2}{27}R^{3}+ \frac{5}{3}RR_{\mu\nu}R^{\mu\nu} -\right.\nn\\
&& -\frac{13}{3}R_{\mu\nu}R^{\mu}{}_{\rho}R^{\nu\rho} \left.\left. +23R_{\mu\nu}R_{\rho\sigma}C^{\mu\rho\nu\sigma}\vphantom{\frac{3}{2}}\right\}\right]+ \mathcal{O}\left( C^2, {\rm GB}_0 \right)\ .
\end{eqnarray}
The simplification is not as radical as in the Ricci--flat case (\ref{grav action R linear}) but still one can straightforwardly find the corresponding equations of motion (the most relevant variation, $\delta C_{\mu\nu\rho\sigma}$, is given in (\ref{varW})):
\bea\label{Conformal EOM}
E^{\alpha\beta}&=&\frac{\Lambda^4 \mu_0}{2}g^{\alpha\beta} + \Lambda^2 \mu_1 \left( R^{\alpha\beta}-\frac{1}{2}g^{\alpha\beta}R \right) + \nn \\
&&+\frac{\mu_3}{5\Lambda^2}\left[ \frac{203}{6}g^{\alpha\beta}R^{\mu\nu}R_{\mu}{}^{\rho}R_{\nu\rho}+\frac{71}{2}R^{\alpha\beta}R^{\mu\nu}R_{\mu\nu}- 105R^{\alpha\mu}R^{\beta\nu}R_{\mu\nu}- \right.\nn\\ &&-\frac{100}{3}g^{\alpha\beta}R^{\mu\nu}R_{\mu\nu}R+81R^{\alpha\mu}R^{\beta}{}_{\mu}R-\frac{163}{6}R^{\alpha\beta}R^{2}+\frac{397}{54}g^{\alpha\beta}R^{3}-\nn\\ &&-\frac{79}{6}RR{}^{\alpha\beta;\mu}{}_{\mu}-\frac{2}{3}R^{\alpha\beta}R{}^{;\mu}{}_{\mu}+\frac{55}{36}g^{\alpha\beta}RR{}^{;\mu}{}_{\mu}- \frac{10}{3}R^{\alpha\beta;\mu}R_{;\mu}-\nn\\ &&-\frac{11}{36}g^{\alpha\beta}R^{;\mu}R_{;\mu}+23R^{\alpha\beta;\mu\nu}R_{\mu\nu}+\frac{29}{6}g^{\alpha\beta}R^{\mu\nu}R_{;\mu\nu}+\nn\\ &&+36R^{(\alpha\mu}R^{\beta)}{}_{\mu}{}^{;\nu}{}_{\nu}+g^{\alpha\beta}R{}^{;\mu}{}_{\mu}{}^{\nu}{}_{\nu}- 36R^{(\alpha\mu;\beta)\nu}R_{\mu\nu}-\nn\\ &&-23R^{\alpha\mu;\nu}R^{\beta}{}_{\nu;\mu}+36R^{\alpha\mu;\nu}R^{\beta}{}_{\mu;\nu}-11g^{\alpha\beta}R^{\mu\nu}R_{\mu\nu}{}^{;\rho}{}_{\rho}+\nn\\ &&+18g^{\alpha\beta}R^{\mu\nu;\rho}R_{\mu\rho;\nu}-11g^{\alpha\beta}R^{\mu\nu}{}^{;\rho}R_{\mu\nu;\rho}- 23R^{\mu\nu;(\alpha}R^{\beta)}{}_{\mu;\nu}+\nn\\ &&+\frac{25}{3}R^{(\alpha\mu;\beta)}R_{;\mu}+11R^{\mu\nu;(\alpha\beta)}R_{\mu\nu}- 13R^{(\alpha\mu;\nu}R_{\mu\nu}{}^{;\beta)}+\nn\\
&&+11R^{\mu\nu;\alpha}R_{\mu\nu}{}^{;\beta}-\frac{43}{36}R^{;\alpha}R^{;\beta}- \frac{44}{3}R^{(\alpha\mu}R^{;\beta)}{}_{\mu}+\nn\\
&&+\left. \frac{91}{18}R^{;\alpha\beta}R-R^{;\mu}{}_{\mu}{}^{\alpha\beta}\right].
\eea

One of the most important conformally flat backgrounds is the cosmological Friedmann--Lema\^ i{}tre--Robertson--Walker (FLRW) spacetime (which is conformally flat for any value of the FLRW topology index $k=-1,0,+1$ \cite{Iihoshi:2007uz})
\begin{eqnarray}\label{FRW}
ds^2 = a(t)^{2}\left( -dt^{2}+\frac{dr^{2}}{1-kr^2}+r^{2}d\Omega^{2}\right)\ .
\end{eqnarray}
Plugging this into (\ref{Conformal EOM}) one obtains the following non-zero components for the tensor of equations of motion $E^{\alpha}{}_\beta$ (only the diagonal components are non-vanishing):
\begin{eqnarray}\label{FRW EOM}
E^{t}{}_{t}&=&\frac{1}{2}\Lambda^4\mu_0-3\Lambda^2\mu_1\left(\frac{a'^{2}}{a^{4}}+\frac{k}{a^{2}}\right)+\frac{\mu_3}{\Lambda^2}
\left( \frac{137a'^{6}}{a^{12}}+\frac{147ka'^{4}}{5a^{10}}+\frac{16a''^{3}}{a^{9}}+\frac{9a^{(3)2}}{5a^{8}}-\right.\nn\\
&&-\frac{21ka''^{2}}{5a^{8}}-\frac{360a'^{4}a''}{a^{11}}+\frac{132a^{(3)}a'^{3}}{a^{10}}+\frac{150a'^{2}a''^{2}}{a^{10}}-\frac{144a^{(4)}a'^{2}}{5a^{9}}-\frac{168ka'^{2}a''}{5a^{9}}+\nn\\
&&\left.+\frac{18a^{(5)}a'}{5a^{8}}-\frac{18a^{(4)}a''}{5a^{8}}+\frac{42ka^{(3)}a'}{5a^{8}}-\frac{48a^{(3)}a'a''}{a^{9}}\right)\nn\\
{\rm and}&&\nn\\
E^{r}{}_{r}&=&E^{\theta}{}_{\theta}=E^{\phi}{}_{\phi}=\frac{1}{2}\Lambda^4\mu_0+\Lambda^2\mu_1\left(\frac{a'^{2}}{a^{4}}-2\frac{a''}{a^{3}}-\frac{k}{a^{2}}\right)+\nn\\
&&+\frac{\mu_3}{\Lambda^2}\left(-\frac{411a'^{6}}{a^{12}}-\frac{343ka'^{4}}{5a^{10}}+\frac{68a''^{3}}{a^{9}}-\frac{19a^{(3)2}}{a^{8}}-\frac{77ka''^{2}}{5a^{8}}+\frac{6a^{(6)}}{5a^{7}}+\frac{14ka^{(4)}}{5a^{7}}+\right.\nn\\&&+\frac{1234a'^{4}a''}{a^{11}}-\frac{428a^{(3)}a'^{3}}{a^{10}}-\frac{830a'^{2}a''^{2}}{a^{10}}+\frac{508a^{(4)}a'^{2}}{5a^{9}}+\frac{532ka'^{2}a''}{5a^{9}}-\frac{78a^{(5)}a'}{5a^{8}}-\nn\\&&\left.-\frac{146a^{(4)}a''}{5a^{8}}-\frac{126ka^{(3)}a'}{5a^{8}}+\frac{328a^{(3)}a'a''}{a^{9}}\right).
\end{eqnarray}\con

One notices that the EOM evaluated on FLRW background do not depend at all on the coefficient $\mu_2$. This is actually true on any conformal background. The reason for this is that the term in the expansion of the spectral action proportional to $\mu_2$ is precisely with four derivatives and as found in \cite{Connes:1990qp,Connes:1994yd} it is exactly conformally invariant $\sqrt{g} C^2$ term, see (\ref{grav action W}). Hence there is no contribution of the first variation of it on the conformal background. From this one derives that conformally flat backgrounds are exact solutions in the same way (we mean that exactly the same form of the source is needed) as in two-derivative Einstein theory with a cosmological constant, when the expansion to the order of $a_4$ is retained. However, as seen from above equations, the inclusion of the next term in the expansion -- the $a_6$ coefficient changes this conclusion and we get many non-zero terms proportional to $\mu_3$ in EOM. This means that we cannot rely on cosmological solutions of Einstein-Hilbert theory possibly with a cosmological constant and the set of Eqs. (\ref{FRW EOM}) has to be solved anew. For example, the question whether one can find some well-behaved solution for the scale factor $a(t)$ for some reasonable cosmological energy-momentum tensor requires further study.

\hypersetup{bookmarksdepth=-2}
\pdfbookmark[1]{4 Some quantum properties of the model}{name8}
\section{Some quantum properties of the model}
\label{quantum_beta}
\hypersetup{bookmarksdepth}

As we saw in the previous section, using just the classical analysis of the derivative expansion of the spectral action it is difficult to see if there is anything special about it. As a next step, it is very important to check whether the spectral action is preferred on quantum level. In this section we make some initial effort in this direction, postponing the detailed study for the future research.

To proceed, one should quantize the higher derivative theory given for example by the expansion of the spectral action up to the $a_6$ coefficient. (The quantization of the action up to $a_4$ was already considered by Stelle in \cite{stelle1} since the model resulting from this level of expansion of the spectral action is a four-derivative theory only with $C^2$ and $\rm GB$ terms (without $R^2$ term), with Einstein-Hilbert term $R$ and a non-zero cosmological constant.) For the covariant quantization one can use the method presented in \cite{BOS} or one may desire to use Batalin--Vilkovisky formalism from \cite{Batalin1,Batalin2} to have better control over remaining BRST symmetry of the quantized theory. Since the special attention was paid to conformal backgrounds in previous studies (in particular to 4-dimensional product manifolds of the type $S^1\times S^3$) in \cite{Connes:1990qp,Connes:1994yd,Chamseddine:2008zj}, it seems natural to investigate the quantum stability of perturbations around these backgrounds. In some minimal sense one should check the positive-definiteness of the quadratic operator governing the dynamics of  small quantum perturbations around a conformal background. This issue is tightly related to the positivity of beta functions in front of $R^2$ and $C^2$ invariants in the form of the one-loop divergent effective action in the theory. In Euclidean framework both these curvature invariants are positive-definite. In a bigger generality one could consider the whole system of beta functions for the quantum theory, not only in front of dimensionless (in $d=4$) terms $R^2$ and $C^2$, but also the beta function of the cosmological constant $\beta_{\rm cc}$ and the beta function $\beta_G$ of the Newton's constant coupling $G_N$. This last beta function is defined as the divergent coefficient in front of Ricci scalar term $R$ in the one-loop divergent effective action. Actually, for the last two beta functions $\beta_{\rm cc}$ and $\beta_G$ we know the answer in general higher derivative theories. The easier computation of the beta function $\beta_{\rm cc}$ was first done in \cite{shapiro3}, while the more involved computation of $\beta_G$ involving contributions from generalized Gauss-Bonnet terms was achieved in \cite{RGsuperren}. The analysis presented in \cite{shapiro3} and \cite{RGsuperren} is generally valid on any background spacetime but obviously very easily we can restrict it to a preferred conformal background or even to a particular example of $S^1\times S^3$ manifold.

First, one can understand that terms cubic in curvatures do not contribute to  the beta function $\beta_{\rm cc}$.
This statement is based on the argumentation presented in \cite{superrenfin,universality,RGsuperren}. One may say in simple words that all the terms in the ``potential'' depending on the curvature $V({\cal R})$ do not influence at all the beta function of the cosmological constant in the theory. However, it is expected that they will contribute to two beta functions of dimensionless couplings $\beta_{R^2}$ and $\beta_{C^2}$ as well to $\beta_G$. Actually, the computation of the two remaining beta functions $\beta_{R^2}$ and $\beta_{C^2}$ is one of the very important goals of the extension of the project, which we plan to address in the nearest future. When one checks the actual expression for the beta function $\beta_{\rm cc}$, one sees almost no speciality of the theory based on the action (\ref{grav action 0}). Since this beta function is completely insensitive to terms cubic in curvature in (\ref{grav action 0}), see below, we can concentrate only on terms quadratic in curvature. These terms in the action are, of course, very important for defining the kinetic operator and hence ensuing propagator for gravitational quantum fluctuations around flat spacetime background. If we could see any extraordinary behavior of the system of beta functions here, we must emphasize that this would not be a virtue of spectral action approach only since the latter constrains tightly also the numerical coefficients in front of cubic terms, but $\beta_{\rm cc}$ does not depend on them. Instead the special behavior could be associated to hypothetic structural relations between the terms quadratic in curvature describing kinetic part of the theory. For analysis of the beta function we can use either the Weyl--dominated basis (\ref{grav action W}) or the ($R,C,{\rm GB}$)-basis (\ref{grav action GB}). Following discussions in \cite{shapiro3,RGsuperren}, we note that for the beta function we need to focus on the coefficients in front of the terms with respectively two derivatives, four derivatives and six derivatives being also quadratic in curvatures. Towards this end, let us write (\ref{grav action W}) or (\ref{grav action GB}) in the form (\ref{RCGB})
\bea
S_{\rm grav} &=& \!\int\!d^4x\sqrt{g}\left(c_{-2}+c_{-1}R+c_{0}^R R^2+c_{0}^C C_{\mu\nu\rho\sigma}C^{\mu\nu\rho\sigma}+c_{1}^R R\square R+c_{1}^C C_{\mu\nu\rho\sigma}\square C^{\mu\nu\rho\sigma}\right)+\nn\\
&&+{\cal O}\left({\cal R}^3\right),
\label{action_shapiro}
\eea
where
\begin{equation}
c_{-1}=-\Lambda^2\mu_1,\quad c_{0}^R=0,\quad c_{0}^C=-\frac{3}{2}\mu_2,\quad c_{1}^R=-\frac{1}{10}\frac{\mu_3}{\Lambda^2},\quad c_{1}^C=-\frac32\frac{\mu_3}{\Lambda^2}.
\label{omega_assign}
\end{equation}
We notice right away that the beta function does not depend on the cosmological constant term $c_{-2}$ (because there are no derivatives in this term -- and the difference in energy dimensionalities of this
term compared to the coefficients $c_1^R$ and $c_1^C$ is bigger than the number of dimensions $d=4$), neither on $c_{0}^{\rm GB}$ (because this is a topological term in $d=4$), nor on $c_{1}^{\rm GB}$ (because this term can be re-written in terms that are cubic in curvatures \eqref{GB1_Int}). The result for the beta function $\beta_{\rm cc}$ from \cite{shapiro3} (where we have to take $N=1$ corresponding to the theory with six derivatives) reads
\begin{equation}
\beta_{{\rm cc}}=-\frac{1}{2(4\pi)^{2}}\left[c_{-1}\left(\frac{1}{3}\frac{1}{c_{1}^R}-5\frac{1}{c_{1}^C}\right)+\left(\frac{c_{0}^R}{c_{1}^R}\right)^{2}+
5\left(\frac{c_{0}^C}{c_{1}^C}\right)^{2}\right].
\label{beta_cc}
\end{equation}
Plugging into this (\ref{omega_assign}), one finds
\begin{equation}
\beta_{{\rm cc}}=-\frac{1}{(4\pi)^{2}}\frac{5}{2}\Lambda^{4}\left(\frac{\mu_{2}}{\mu_{3}}\right)^{2}.
\label{beta_cc_final}
\end{equation}
We comment on some simplification which occurred above. First, the second term in the square bracket in \eqref{beta_cc} is not present since $c_{0}^R=0$ as this was discussed to be a feature of the spectral action to the order $a_4$ in expansion. However, the vanishing of the first term proportional  to $c_{-1}$ is a genuine feature of the coefficients appearing in the expansion to the level of $a_6$. The relation between $c_{1}^R$ and $c_{1}^C$ (that is $c_{1}^C=15c_{1}^R$) is dictated by spectral action approach, but as we emphasized above cubic terms in $V(\cal R)$ do not participate, so right now we cannot judge whether this is a mere numerical coincidence or some deeper fact related to the roots of spectral action and non-commutative geometry approaches. We put importance to the fact that this relation holds independently of the value of the dimensionful cut-off parameter $\Lambda$ as well as of the arbitrary and adjustable value of the dimensionless coefficient $\mu_3$. However, from the field theory point of view, there is not much of importance of this observation, since the total beta function is non-zero. As far as we know there does not exist any clear interpretation of the fact that the final expression for the cosmological constant beta function is independent of the value of $c_{-1}$ coupling, which stands in front of the Ricci scalar in the action \eqref{action_shapiro}. The final expression for the beta function \eqref{beta_cc_final} shows that it is always negative-definite and that it depends on the value of the ratio of the coefficients $\mu_2/\mu_3$ only. There is a very little amount of speciality of the quantum behavior of the spectral action.

We also remark that using the analysis presented in \cite{RGsuperren} we cannot unambiguously determine $\beta_G$ since in our model (\ref{grav action GB}) we have other terms cubic in curvature besides the generalized Gauss-Bonnet term $\rm GB_1$, while the analysis of \cite{RGsuperren} was done for the ``minimal'' model with $V({\cal R})=0$.
Some preliminary results indicate that there is no exceptional behavior of the other three beta functions of the theory, that is we do not find any of $\beta_{R^2}$, $\beta_{C^2}$ and $\beta_G$ to be zero or to be always strictly positive though some further analysis is still required.

\hypersetup{bookmarksdepth=-2}
\pdfbookmark[1]{5 Discussion and conclusions}{name9}
\section{Discussion and conclusions}
\label{sec4}
\hypersetup{bookmarksdepth}

In this paper, we studied some classical aspects of a specific higher derivative gravity theory motivated by the spectral action approach. One of the aims of this work was to bring attention of the researchers working in higher derivative gravity to the methods of non-commutative geometry. This goal partly defined the style of the paper - along with the original research, it contains some details (mostly collected in the appendices) known to those who work in non-commutative geometry but mostly unfamiliar to the higher derivative gravity community.
One of the main motivations to consider the spectral action as the basis for the effective higher derivative gravity is the fact that the derivative expansion has a fixed structure within each derivative level, greatly reducing the dimension of the parameter space. This gives hope that the spectral HD gravity might have some special properties compared to the general case. The immediate analysis is difficult due to the very ``bulky'' form of the general expressions for the relevant terms in the expansion of the spectral action. So, the major part of the paper is devoted to deriving the most compact form of the 6-derivative part as well as some equivalent representations, which might be useful for different types of problems. The formulas (\ref{grav action R}), (\ref{grav action W}) and (\ref{grav action GB}) constitute ones of the main technical results of our work.

As we mentioned above several times, the rigidity of the structure of the higher derivative terms gives hope that the theory might possess the features absent in the general case. This hope is somewhat supported by the observation made in the paper with the title suggesting the existence of such special features - ``The Uncanny Precision of the Spectral Action'' \cite{Chamseddine:2008zj}. There it was shown that on the special type of a conformal background, $S^1 \times S^3$, the higher derivative part of the action is identically zero. Our result (\ref{grav action W}) allowed us to study this point in great generality. We showed that the result of \cite{Chamseddine:2008zj} does not hold for a general conformal background, so it is, in some sense, accidental (or signalling that $S^1 \times S^3$ background is in some way special). In particular, the action is not trivial for one of the most physically relevant conformal backgrounds - cosmological spacetimes. To continue the study of the classical gravity based on spectral action, we derived the general equations of motion, as well as their special cases - for Ricci--flat and conformal backgrounds. While the general EOM, do not seem to be particulary simple, in the Riemann/Weyl basis there are serious simplifications. As an (somewhat trivial) application, we explicitly demonstrated that neither Schwarzschild nor cosmological spacetimes are the exact solutions of these equations for the same matter energy-momentum source as this was in standard Einstein gravity.

At this point, it seems that the main conclusion of the classical analysis is that at this level there is nothing much special about the specific values of the parameters fixed by the spectral action. While this appears to be the case, it does not mean that the same should be said in general. The reason is that there is still a chance that the special values of the parameters will be important at the quantum level. In this work, using the example of the cosmological constant beta function, $\beta_{\rm cc}$, we briefly touched upon the possible implications of the spectral action approach on quantum level. But much more detailed study is still needed. So, naturally this should be one of the most urgent next steps in the continuation of this project.

We want to discuss here the issue of the dependence of the spectral action on the order of  expansion. We can now compare results (both classical and quantum) in higher derivative gravitational theories  based on the action given up to $a_4$ and $a_6$ coefficients. One sees that inclusion of terms with higher number of derivatives changes theory quite dramatically. For example, in the domain of classical exact solutions (and their stability properties) we observed a lot of differences between the two models as discussed in Section \ref{EOM}. One may ask whether the inclusion of terms with six derivatives of the metric tensor present in $a_6$  is a small perturbation added to the system. From the field theory viewpoint, this is \emph{not} the case. Classical EOM change their character from forth to sixth order in derivatives and this implies that we have two new families of solutions for each problem. One might think that however, the perturbation by sixth derivative term is small and it modifies the known solutions (from four derivative theory or even from Einstein theory) only by a little. But due to the higher derivative character of modification we see strong differences both in the IR (long wavelengths) as well as in the UV-regime (boundary with quantum microscopic domain). The first regime exhibit differences because of the new families of solutions (like runaway solutions compared to $1/r$ Newtonian potential solution). Whereas in the short distances regime the terms with higher derivative again start to lead and dominate over terms with lower number of derivatives because generally this regime is identified with high energies and then the more derivatives we have in the action or EOM, the higher power of energy or momentum we have in the corresponding solutions (compare this to the discussion of the scaling dimension in \cite{Pinzul:2010ct}). In ordinary field theory it is possible to conceive modifications which are true small perturbations (like a non-derivative interaction in renormalizable scalar field models), however, in gravitational setup we are doomed to consider only higher derivative modification of the Einstein-Hilbert plus cosmological constant action. Such deformations of the standard gravitational theory cannot be considered as perturbative since it is difficult to find a regime in which they are not the dominant ones over the terms with lower number of derivatives (compare also discussion in \cite{Simon1,Simon2}). This remark applies not only to the jump from $a_2$ to $a_4$ but also from $a_4$ to $a_6$ or from $a_6$ to higher orders in the expansion. Therefore the question arises whether we should trust more the results obtained in a higher truncation based on $a_6$ than on $a_4$ and whether the results and conclusions there will not be washed away by consideration of the even more accurate model based on $a_8$ coefficient of the expansion and so on.

One of the possible solutions to this problem is naturally given by the spectral action approach since there are two ingredients which could help us. First one is the presence of the arbitrary energy scale $\Lambda$. Thanks to this, we can treat the terms in the expansion of the spectral action as terms in an asymptotic series in $\Lambda^{-1}$ variable. Then despite that numerical coefficients of higher derivative terms are finite (not infinitesimal!) numbers we can make them perturbative by considering $\Lambda$ very big compared to other energy scales present in the system (for example comparing to electroweak symmetry breaking scale in the Standard Model $E\approx216\,{\rm GeV}$). Strictly speaking the impact of higher derivatives is perturbative only when the scale $\Lambda$ is sent to infinity. Another source for justification of the perturbative treatment comes with the coefficients $\mu_2$ and $\mu_3$. They depend on the precise form of the cut-off function as described in Section \ref{sec2}. However, from physical requirements of having a good decoupling of high energy modes, the cut-off profile should be very close to flat near zero. This means that the $\mu_3$ coefficient should be very small. And this provides an additional suppression of the higher derivative terms and allows to treat their impact on classical exact solutions as small. However, one can see that assumption $\mu_3\ll\mu_2\ll1$ blows up the expression for the beta function in \eqref{beta_cc_final}. Then to render it finite one must enter into a game of playing with three parameters $\Lambda$, $\mu_2$ and $\mu_3$, which is significantly more complicated and will be discussed elsewhere.

Actually the problems with dependence on the level of expansion are much deeper on the quantum level. To have a renormalizable model of quantum gravity one must treat higher derivatives as the leading and dominant terms, not as perturbatively small additions to perturbatively non-renormalizable Einstein-Hilbert gravitational action. When one does this, one indeed finds that the model based on the expansion up to $a_6$ coefficient is renormalizable. (The model with $a_4$ coefficient is formally non-renormalizable because it does not contain in the action the term with $R^2$ but the stronger reason is the presence of conformal anomaly in this model \cite{confreview}.) Actually, using the definitions in \cite{BOS,superrenfin} this is a three-loop super-renormalizable model of QG, meaning that the last divergences are met on the level of three-loop computation, while from the forth loop and upwards the theory is completely UV-finite. Similarly, when we discuss the form of one-loop beta functions (related to perturbative UV-divergences) we assume that the terms giving rise to the UV behavior of the propagator are from the terms in the action with the highest number of derivatives. Not assuming this non-perturbative character of higher derivative terms would immediately spoil super-renormalizability and renormalizability of the model \footnote{If one includes the effects of higher derivative terms only as vertices of the theory, while keeps at the same time propagator derived from terms with less derivatives, then new perturbative UV-divergences pop out. These divergences contain more derivatives, more even than there are in
originally added higher derivative terms. Hence such theory is perturbatively non-renormalizable.}. The UV behavior of the propagator for gravitational perturbations is the crucial thing for the discussion of any UV properties of the theory. For any local higher derivative theory the procedure of finding the UV behavior of the propagator consists of looking for the terms in the action, which are quadratic in curvatures and with the highest, but finite, number of derivatives on the metric tensor. These terms shape the ultra-violet form of the kinetic operator for quantum fluctuations. One understands that the beta functions in the model based on $a_6$ are different from the ones in the model based on $a_4$ and there does not exist any limit which makes the two match, which is obvious from the explicit formulas for beta functions in \cite{RGsuperren}. Therefore one cannot treat the six-derivative terms in $a_6$ as quantum perturbations in any sense. Moreover, for the derivation of beta functions of the theory the terms with six derivatives are the leading ones in the UV and hence cannot be considered small in this regime.

Analogous problem we meet when we search for the perturbative spectrum of fluctuations. For definiteness we can study this spectrum around flat spacetime background. In order to find poles of the propagator in respective sectors of spin-2 and spin-0 fluctuations, one needs all set of terms which are quadratic in curvatures and the term linear in Ricci scalar and cosmological constant term. Inevitably in higher derivative theories we are faced with the problem of perturbative ghosts in the spectrum. These virtual states have negative sign of the kinetic term (for tachyons they have negative real part of the mass square parameters), hence they endanger perturbative unitarity of the theory. For example, the optical theorem for on-shell scattering amplitudes does not hold anymore. These are undesirable states and they should be eradicated from the theory by all means. For removing them (or their effects on observable predictions of HD theories) various approaches have been introduced: Lee--Wick prescription \cite{LW1,LW2}, fakeons \cite{Ans1,Ans2,Ans3}, disappearance of unstable perturbations on non-trivial backgrounds \cite{shapiro5}. However, none of the proposals seems to be completely satisfactory. Of course, one can always argue that the full spectral action will give rise to also non-perturbatively unitary higher derivative quantum field theory of gravitational interactions and blame the apparent non-unitarity as the result of truncation of the spectral action to some finite-order HD models, but to make this statement precise much more of a very non-trivial analysis should be done.

In our case to find poles of the propagator (or equivalently zeros of the kinetic operator governing dynamics of quantum perturbations), one needs to know all the coefficients in front of the terms quadratic and linear in curvatures. This is in distinction to the computation of UV-divergences where we needed only the coefficients of few terms with the highest number of derivatives in UV (in $d=4$ we need coefficients of terms with highest number of derivatives and the ones with two and four less derivatives only). The reason for this is that the beta functions are the UV issue while the spectrum is the problem at all energy scales. The necessary information is given in \eqref{omega_assign} and the value of the cosmological constant coupling, $c_{-2}=\Lambda^4\mu_0$. Once again we do not have any contribution from terms which are cubic in curvatures (this is true for flat spacetime propagator). Another problem is that for the theory with cosmological constant $c_{-2}\neq0$ flat spacetime is not an on-shell background. (It does not satisfy vacuum gravitational EOM with the cosmological constant term and none energy-momentum source of matter origin on the RHS of gravitational EOM.) Then we cannot consider quantum dynamics of fluctuations in the WKB approximation and the analysis of the propagator around flat background is purely academic. However, mathematically, as a demonstration, one can neglect this obstacle and continue with the analysis. The zeros of the kinetic operator are zeros of the respective polynomials in $k^2$ variable in momentum space, in two gauge-invariant sectors of spin-2 (related to the terms quadratic in Weyl tensor) and spin-0 (related to the terms quadratic in Ricci scalar) fluctuations. These zeros describe the mass square parameters of the modes. In HD models we always meet ghosts \cite{shapiro3} as the consequence of UV-improved behavior of the theory compared to two-derivative theories. In our case, the theory is based on the action functional given in \eqref{grav action GB} and we have that both polynomials are of the third order in $k^2$ variable. This means that in each sector we expect three (possibly some are equal), in general, complex roots describing mass square parameters. For the precise values we need to solve cubic equations. We will not do this here, but we will comment on the general features of these solutions. The exact values depend on the numerical values of $\mu_k$ parameters (for $k=0,1,2,3$) and on $c$'s in \eqref{omega_assign}. There are two possibilities: the three roots come in a form of one complex pair (of two complex conjugate roots) and one real root or all three roots are real. The former case is well known and then the pair is called a pair of Lee-Wick particles. They have quite peculiar properties similar a bit to a couple of unstable particles in standard field theories \cite{LWmod1,LWmod2}. Therefore the model with six derivatives may realize the scenario of Lee-Wick quantum gravity (it was impossible to have a pair of complex ghosts in four-derivative theories).

Here, one can also ask the question how stable is the position of poles of the propagator against inclusion of higher terms in the expansion of the spectral action. Again the situation is quite delicate but not as dramatic as for beta functions (where we had discontinuous jumps when we increased the order of the expansion). Because the higher degree polynomials have more solutions on the complex plane and the coefficients of the terms with the highest power exponent on $k^2$ variable are highly suppressed by the scale $\Lambda$, the new roots always come in pairs from the point at complex infinity and the picture (or position on the complex plane) of the other roots is only slightly modified.  This pair of new zeros moves smoothly when the value of the $\Lambda$ parameter is changed from infinity, so the change in the set of zeros is continuous. For example, if we find that the theory based on the spectral action up to the coefficient $a_6$ is a model of Lee-Wick quantum gravity (for some definite values of $\Lambda$, $\mu_0$, $\mu_1$, $\mu_2$, $\mu_3$), then it is likely that this feature of having addditional particles beside the real graviton only in complex conjugate pairs, will be preserved for higher orders in truncation of the spectral action, provided also that the value of the $\Lambda$ parameter is large (then the number of these LW pairs will increase). Hence the Lee-Wick characteristics of models of quantum gravity is quite stable.

One might ask many reasonable questions about what would happen if we had at our disposal the full re-summed spectral action, i.e. if we would have a control over the non-perturbative form of the spectral action. Some of these questions are: What would be the exact classical solutions? Will the quantum theory be eventually unitary, renormalizable, or even UV-finite? To what extent solutions or beta functions based on truncated action reflect the situation in the full theory? Can they be treated as subsequent approximations in some perturbative scheme? We do not have even tentative answers to these important questions and due to  technical reasons we must deal with the expansion of the spectral action in number of derivatives. One should expect some very non-trivial UV properties of the full spectral action \cite{Kurkov:2013kfa}. It is plausible to think that the full theory may take a form of some non-local model of QG as discussed in \cite{nlrev}. And then the expansion that we are performing parallels the limiting method of approaching non-local models by some higher derivative models. Therefore,  with such a perspective the questions of exact solutions and of beta functions acquire completely new answers in full re-summed models. For example for beta functions, we must not look into ratios like $\mu_{n-1}/\mu_n$ (cf. \eqref{beta_cc_final}), but into the limit of these ratios when $n$ is sent to infinity. Then this changes the philosophy and we must instead ask questions about convergence radius of the analytic function given by a formal series $\sum_{n=1}^\infty \mu_n z^n$. Even if we know that formally the term with the highest number of derivatives in such expansion does not exist (firstly, because it is formally with $n=\infty$, secondly because of its coefficient vanishing as $\lim_{n\to\infty} \mu_n=0$), we can still in some sense talk about the beta function in non-local theory which is defined by the convergence radius above. Perhaps, in a similar sense we can talk and define non-perturbative beta functions of couplings in full spectral action. It remains to be seen what is the full analytic structure of the theory based on the full re-summed spectral action and whether this can be mapped to some non-local models of quantum gravity.

\appendix
\setcounter{equation}{0}
\pdfbookmark[1]{A Conventions and useful formulas}{name10}
\hypersetup{bookmarksdepth=-2}
\section{Conventions and useful formulas}
\label{apA}
\hypersetup{bookmarksdepth}

Here we collect some notations and standard formulas used in the main text.

Our conventions for symmetrization and anti-symmetrization of indices are the following: $(\mu_1\cdots\mu_n)$ and $[\mu_1\cdots\mu_n]$ mean respectively complete symmetrization and anti-symmetrization with respect to the indices $\mu_1$ to $\mu_n$, with the proper symmetry factor (that is $1/n!$). In the situation, where there are various operations nested on the same group of indices, the bracket $[\![\cdots]\!]$ will be used for not confusing which pair of the indices is being anti-symmetrized in the second turn (see the appendix \ref{AppEOM}).

Covariant derivatives on the geometric objects (trivial from the point of view of bundle space structure) we denote either in the standard semicolon (;) postfix (GR) notation or in a operatorial prefix notation with the symbols of nabla ($\nabla$), which is however a more frequent choice in field theory.

In the Euclidean signature (used mostly through the text of the article) we choose signature of the metric tensor to be all pluses, and in Minkowski (analytically continued) case we take the time as the first coordinate and choose the signature of the metric to be $(+,-,-,-)$ in four spacetime dimensions. \con

Our convention for overall signs of  Riemann tensor, Ricci tensor and Ricci scalar takes respectively the following forms:
\bea
R_{\mu\nu}{}^\rho{}_{\sigma}V^{\sigma} &=&-\left[\nabla_{\mu},\nabla_{\nu}\right]V^{\rho}\,, \label{sriem}\\
R_{\mu\nu} &=&-g^{\rho\sigma}R_{\mu\rho\nu\sigma}\,,\label{sric}\\
R &=& g^{\mu\nu}R_{\mu\nu} \label{signrs}\ .
\eea
Note the non-standard sign in the definition of the Riemann and Ricci tensors (opposite, for example, to conventions of Landau-Lifshitz \cite{Landau:1980}). This choice is made to agree with the notations used in the literature on the heat kernel expansion \cite{Gilkey:1995mj,Gilkey:2004dm} and on non-commutative geometry \cite{Chamseddine:1996rw,Chamseddine:1996zu,Chamseddine:2008zj}. Using this definition, the commutator of the covariant derivatives acting on a general tensor $T^{\alpha_{1}\ldots\alpha_{m}}{}_{\beta_{1}\ldots\beta_{n}}$ (with $m$ contravariant indices  and $n$ covariant ones) can be written as,
\bea\label{Commut. Cov. Derv.Gen. Tensor.}
[\nabla_{\mu},\nabla_{\nu}]T^{\alpha_{1}\ldots\alpha_{m}}{}_{\beta_{1}\ldots\beta_{n}} &=& -\sum_{i=1}^{m}R_{\mu\nu}{}^{\alpha_{i}}{}_{\sigma}T^{\alpha_{1}\ldots\alpha_{i-1}\sigma\alpha_{i+1}\ldots\alpha_{m}}{}_{\beta_{1}\ldots\beta_{n}}-\nn\\ &&-\sum_{i=1}^{n}R_{\mu\nu\beta_{i}}{}^{\sigma}T^{\alpha_{1}\ldots\alpha_{m}}{}_{\beta_{1}\ldots\beta_{i-1}\sigma\beta_{i+1}\ldots\beta_{n}} \ .
\eea

We remind that Bianchi identities for the Riemann tensor are expressed as
\be
R_{\mu[\nu\rho\sigma]}=0
\ee
and
\be
R_{\mu\nu[\rho\sigma;\kappa]}=0
\ee
with the names of respectively the first and the second identity. Contracting the second Bianchi identity, we get the singly contracted second Bianchi identity
\bea\label{Contrct. Sec. Bianchi 1}
R^{\mu}{}_{\nu\rho\sigma;\mu}=2R_{\nu[\rho;\sigma]}=R_{\nu\rho;\sigma}-R_{\nu\sigma;\rho}\ .
\eea
Contracting one more time, will lead us to the doubly contracted second Bianchi identity:
\bea\label{Contrct. Sec. Bianchi 2_Ric}
R^{\mu}{}_{\nu;\mu}=\frac{1}{2}R_{;\nu}\ .
\eea

The standard expression for the Weyl (conformal) tensor in $d$ dimensions takes the following form,
\bea\label{Weyl}
C_{\mu\nu\rho\sigma}&=&R_{\mu\nu\rho\sigma}+\frac{4}{d-2}g_{[\mu[\![\rho}R_{\sigma]\!]\nu]}-\frac{2}{(d-2)(d-1)}g_{\mu[\rho}g_{\sigma]\nu}R\ .
\eea
One can easily find the following useful expressions:
\bea\label{Weyl Weyl}
C_{\mu\nu\rho\sigma} C^{\mu\nu\rho\sigma}&=&R_{\mu\nu\rho\sigma} R^{\mu\nu\rho\sigma}-\frac{4}{d-2}R_{\mu\nu} R^{\mu\nu}+\frac{2}{(d-2)(d-1)}R^2
\eea
and with one power of the covariant box (covariant d'Alembertian) operator $\square=g^{\mu\nu}\nabla_\mu\nabla_\nu$, inserted:
\bea\label{Weyl box Weyl}
C_{\mu\nu\rho\sigma}\square C^{\mu\nu\rho\sigma}&=&R_{\mu\nu\rho\sigma}\square R^{\mu\nu\rho\sigma}-\frac{4}{d-2}R_{\mu\nu}\square R^{\mu\nu}+\frac{2}{(d-2)(d-1)}R\square R\,.
\eea
Actually, the above formula is valid for any power (or even an analytic function) of the $\square$ operator since it  is a spectator in the derivation. One can notice very big similarity in the structure and coefficients of the corresponding terms between formulas \eqref{Weyl}, \eqref{Weyl Weyl} and \eqref{Weyl box Weyl}. This is not an accidental coincidence and is due to the complete tracelessness property of the Weyl tensor in any dimension. The match would be perfect, if we used the Landau-Lifshitz convention for the overall sign of the Ricci tensor (opposite to the one accepted in \eqref{sric}).

The Gauss-Bonnet scalar is defined by
\bea\label{GB0}
{\rm GB}_0={\rm GB}=R_{\mu\nu\rho\sigma}R^{\mu\nu\rho\sigma}-4R_{\mu\nu}R^{\mu\nu}+R^{2} \ ,
\eea
while its generalization containing $2N+4$ derivatives is given by
\bea
{\rm GB}_{N}=R_{\mu\nu\rho\sigma}\square^N R^{\mu\nu\rho\sigma}-4R_{\mu\nu}\square^N R^{\mu\nu}+R\square^N R\label{GB1} \ .
\eea
In the main text we use the generalized Gauss-Bonet term with $N=1$:
\bea\label{GB}
{\rm GB}_1 := R_{\mu\nu\rho\sigma}\square R^{\mu\nu\rho\sigma} - 4R_{\mu\nu}\square R^{\mu\nu} + R\square R \ .
\eea
While for $N=0$, ${\rm GB}_0={\rm GB}$ is the standard Gauss-Bonnet term, which is topological in the 4-dimensional case (and related there to the Euler invariant), for $N \geqslant 1$ it is not topological anymore but it can be transformed to the form
\be
\mathcal{O}({\cal R}^3) + \nabla_\mu K^\mu \ ,
\nonumber
\ee
where ${\cal R}^3$ stands for different cubic invariants in curvature (see the formula (\ref{GB1_Int}) and the footnote \ref{R_terms}) and $K^\mu$ is a vector field constructed from curvatures and their covariant derivatives, so the last term above is a total derivative.  We remark that since $\rm GB$ is topological in $d=4$ it does not contribute to classical EOM (obtained by the first variation of the action). Away from the case of $d=4$ or for $N\geqslant1$ the generalized Gauss-Bonnet term in the action has an impact on EOM.

\pdfbookmark[1]{B Lichnerowicz formula}{name11}
\hypersetup{bookmarksdepth=-2}
\section{Lichnerowicz formula}\label{App:Lichnerowicz}
\hypersetup{bookmarksdepth}

Here we derive the formula (\ref{Lichnerowicz}) both, to make the presentation more accessible and self-contained, and to introduce our notations and conventions regarding spinors and Dirac operator.

We define the algebra of the flat gamma matrices with the minus sign:
\bea
\{\gamma_a , \gamma_b\} = -2\delta_{ab}\ ,
\label{DiracA}
\eea
where $\delta_{ab}$ is the metric of the flat $d$-dimensional Euclidean space in Cartesian coordinates.
It is well known that $\Sigma_{ab}:=\frac{1}{2}\gamma_{ab}$, where $\gamma_{ab}:=\frac{1}{2}[\gamma_a , \gamma_b]$, are the generators of the Euclidean version of the Lorentz symmetry, i.e. of the orthogonal group $SO(d)$ satisfying the following commutation relations
\be\label{SO4}
[\Sigma_{ab} , \Sigma_{cd}]=-\delta_{ac}\Sigma_{db}+\delta_{bc}\Sigma_{da}+\delta_{ad}\Sigma_{cb}-\delta_{bd}\Sigma_{ca}=-4\delta_{[a[\![c}\Sigma_{d]\!]b]} \ .
\ee
This choice of the Clifford algebra \eqref{DiracA} forces us to use the following as the action for the standard Dirac spinor (two-derivative theory) on flat Euclidean space background:
\be
S_{D}=\!\int\!d^dx \bar{\psi}(-i\partial\!\!\!\slash-m)\psi=\!\int\!d^dx \bar{\psi}(-i\delta^{ab}\gamma_a\partial_b-m)\psi\,.
\ee
Introducing the tetrads associated with the metric $g_{\mu\nu}$ of the curved space
\be
\delta_{ab}e^a_\mu e^b_\nu = g_{\mu\nu}\ ,\ g^{\mu\nu}e^a_\mu e^b_\nu = \delta^{ab}
\ee
we can define the curved gamma matrices by
\be
\gamma_\mu = \gamma_a e^a_\mu
\ee
and they satisfy the anti-commutation relation
\be \{\gamma_\mu , \gamma_\nu\} = -2g_{\mu\nu}\ .
\ee
In passing, we note that the small Greek letters we use for curved (world) space indices, while the small Latin letters we use exclusively for denoting flat (tangent) space indices. In the former space we use the curved metric $g^{\mu\nu}$ to raise world indices, while in the latter flat space we use the Kronecker delta tensor $\delta^{ab}$ to do the corresponding operation on flat indices.

Using these notations the standard Dirac operator is given by
\be\label{Dirac}
\mathrm{D} = \gamma^\mu (\partial_\mu - \omega_\mu)=: \gamma^\mu \nabla^\omega_\mu\ ,
\ee
where $\omega_\mu = \frac{1}{4}\omega^{ab}{}_\mu\gamma_{ab}$ is determined by the requirement that the tetrads are covariantly constant with respect to the covariant derivative $\nabla_\mu^{\omega}$
\be\label{tetrad_0}
\nabla_\mu^{\omega} e^a_\nu = \partial_\mu e^a_\nu + \delta_{bc}\omega^{ab}{}_\mu e^c_\nu - \Gamma^{\rho}{}_{\mu\nu} e^a_\rho = 0 \ .
\ee
Here $\nabla^\omega_\mu = \nabla_\mu - \omega_\mu$ is the \textit{total} covariant derivative, while $\nabla_\mu$ is the usual, Levi-Civita, one.\footnote{This is the same definition as in (\ref{Dirac}), taking into account that acting on a spinor (being in a representation not carrying any Lorentz indices, so not on a gravitino) $\nabla_\mu$ is just a partial derivative $\partial_\mu$.} Since the tetrad (vielbein) is valued both in the tangent as well as curved space (it possesses both types of indices), then the total covariant derivative $\nabla_\mu^{\omega}$ must include connection coefficients from both spaces. In the external space these are given by standard (metric) Christoffel coefficients (and then $\nabla_\mu e^a_\nu=\partial_\mu e^a_\nu- \Gamma^{\rho}{}_{\mu\nu} e^a_\rho$), while in the tangent space this role is played by the $SO(d)$ gauge potentials $\omega^{ab}{}_\mu$.

The main result that allows the direct application of the heat kernel techniques is the Lichnerowicz formula
\be\label{Dirac_square}
\mathrm{D}^2 = -\left(g^{\mu\nu}\nabla^\omega_\mu \nabla^\omega_\nu + \mathbb{E}\right)\ ,\ \ \mathbb{E}:=-\frac{1}{4}R\,\mathds{1}\ ,
\ee
where $R$ is the scalar curvature of the metric $g_{\mu\nu}$. Due to the importance of this formula let us sketch its proof.

We will need several identities:
\begin{itemize}
\item $\nabla^\omega_\mu\gamma^\nu \equiv \partial_\mu \gamma^\nu - [\omega_\mu , \gamma^\nu] = 0$, i.e. $\gamma^\nu$ is covariantly constant with respect to the \textit{total} covariant derivative defined just after (\ref{tetrad_0}). This is an immediate consequence of the analogous statement about the tetrads (\ref{tetrad_0}).
\item The commutator of two total covariant derivatives reads
\be\label{Omega}
-[\nabla^\omega_\mu ,\nabla^\omega_\nu]=\frac{1}{4}R_{\mu\nu}{}^{\rho\sigma}\gamma_{\rho\sigma}=:\Omega_{\mu\nu}\ ,
\ee
which is nothing but the second Cartan equation (after trivially using (\ref{SO4}) or directly commuting the gamma matrices).\footnote{The choice of a sign in (\ref{Omega}) agrees with the convention for the sign of the Riemann tensor stipulated in \eqref{sriem}.}
\item $R_{\mu\nu\rho\sigma}\gamma^{\mu\nu}\gamma^{\rho\sigma}\equiv R_{\mu\nu\rho\sigma}\gamma^{\mu}\gamma^{\nu}\gamma^{\rho}\gamma^{\sigma} = 2R$. This is easily proven noticing that $\gamma^{\nu}\gamma^{\rho}\gamma^{\sigma}=$ $ = \gamma^{[\nu}\gamma^{\rho}\gamma^{\sigma ]} - g^{\nu\rho}\gamma^{\sigma}-g^{\rho\sigma}\gamma^{\nu}+g^{\nu\sigma}\gamma^{\rho}$ and using the first Bianchi identity, $R_{\mu [\nu\rho\sigma ]}=0$.
\end{itemize}
Using these identities it is straightforward to calculate $\mathrm{D}^2$:
\bea
\mathrm{D}^2 &=& \gamma^\mu \nabla^\omega_\mu \gamma^\nu \nabla^\omega_\nu = \gamma^\mu \gamma^\nu \nabla^\omega_\mu \nabla^\omega_\nu = -g^{\mu\nu}\nabla^\omega_\mu \nabla^\omega_\nu + \gamma^{\mu\nu}\nabla^\omega_\mu \nabla^\omega_\nu = \nonumber \\
&=& -g^{\mu\nu}\nabla^\omega_\mu \nabla^\omega_\nu +\frac{1}{8}\gamma^{\mu\nu}R_{\mu\nu}{}^{\rho\sigma}\gamma_{\rho\sigma} \equiv -\left( g^{\mu\nu}\nabla^\omega_\mu \nabla^\omega_\nu - \frac{1}{4} R \right)\ .
\eea

\newpage
\pdfbookmark[1]{C Calculation of the trace}{name12}
\hypersetup{bookmarksdepth=-2}
\section{Calculation of the trace}\label{app_trace}
\hypersetup{bookmarksdepth}

The spectral action is a special case of the following more general expression:
\be\label{trace}
\Tr \chi (P)\,,
\ee
where $\chi$ is some ``more or less'' nice function (the exact meaning of this will be given below) and $P$ is some positive-definite operator on a Hilbert space. In our case, $P=-\mathrm{D}^2$ is represented on the Hilbert space of square-integrable spinors. Because the result for this expression in terms of the heat kernel expansion is one of the main tools in our approach, here we give a detailed (and more or less rigorous) derivation of this expansion. Also, this will hopefully make the paper self-contained.

Let us start by requiring for $\chi=\chi(p)$ to be a piecewise continuous function on $\mathbb{R}^+$ such that
\be
\lim_{p\rightarrow 0^+}\frac{\chi (p)}{p^{a_1}} = b_1\in\mathbb{R}/\{0\} \quad \mbox{ and }\quad \lim_{p\rightarrow +\infty}\frac{\chi (p)}{p^{a_2}} = b_2\in\mathbb{R}/\{0\}\,.
\ee
In other words, we require that the small-$p$ asymptotics (near $p=0$) is given by $b_1 p^{a_1}$ ($\chi(p)=O(p^{a_1})$ for $p\to0$) and similarly the large-$p$ asymptotics (in the $p\to+\infty$ limit) is given by $b_2 p^{a_2}$ ($\chi(p)=O(p^{a_2})$ for $p\to+\infty$). We also demand that $a_2<a_1$.
The interval $(-a_1 ,-a_2)$ is called the fundamental strip of $\chi$. E.g., if $\chi (p)$ is some smooth cut-off function, which at infinity goes to zero faster than any negative degree monomial of $p$ and is of order of $p^0$ when $p\rightarrow 0$, then the fundamental strip is $(0, +\infty)$. Also, let the integral
\be\label{Mellin}
\phi (s) = \int^\infty_0 p^s \chi (p)\frac{dp}{p}
\ee
be convergent when $s$ belongs to the fundamental strip (the function $\phi(s)$ is called a Mellin transform of the function $\chi(p)$), then $\chi (p)$ can be recovered via the inverse Mellin transform:
\be\label{inverseMellin}
\chi (p) = \frac{1}{2\pi i}
\int^{c+i\infty}_{c-i\infty} p^{-s} \phi (s) ds \ ,
\ee
where $c$ is a number which should also belong to the fundamental strip.

Let $P$ be a positive-definite operator and $\chi(p)$ be some cut-off function (i.e. its fundamental strip is $(0, +\infty)$) with $\phi (s)$ being its Mellin transform. Then, using the spectral functional calculus, we can define a function of an operator $P$
\be
\chi (P) = \frac{1}{2\pi i} \int^{c+i\infty}_{c-i\infty} P^{-s} \phi (s) ds \ ,
\ee
where $\phi (s)$ is given by (\ref{Mellin}). Then the functional (or \emph{total}) trace of $\chi (P)$ (\ref{trace}) is given by
\be
\Tr\chi (P) = \frac{1}{2\pi i}\int^{c+i\infty}_{c-i\infty} \zeta_P (s) \phi (s) ds \ ,\label{main}
\ee
where we have introduced the generalized $\zeta$-function:
\be\label{zeta}
\zeta_P (s) := \Tr P^{-s}\ .
\ee
The integration contour in (\ref{main}) can be chosen in such a way that it encircles all the poles of the integrand, that is all the poles are inside the contour and the contour is closed at infinity. It is possible to show that the contribution of the integration over this part of the contour is zero. We will find the poles of $\zeta_P
(s) \phi (s)$ by some indirect method -- using the known results for the heat kernel expansion. The relevance of the heat kernel will become clear after re-writing the zeta-function (\ref{zeta}) in terms of
\be \Tr e^{-tP}\,, \label{hk0}
\ee
which is the object called a trace of heat kernel.

Towards this end, let us use Cahen-Mellin integral (which we will also need later) and its
inverse:
\bea
e^{-p} &=& \frac{1}{2\pi
i}\!\int^{c+i\infty}_{c-i\infty}\! p^{-s} \Gamma (s) ds
\ ,\ \ \ c>0\ ,\ \Re (p) > 0
\eea
with the standard integral definition of the Gamma function $\Gamma(s)$:
\bea
\Gamma (s) &=& \!\int^\infty_0\! x^{s-1} e^{-x} dx \ ,\ \ \Re (s) >0\ .\label{gamma}
\eea
By doing formal change of variable in the last integral, $x\rightarrow tP$, (again using the functional calculus for a positive-definite operator), we have
\be
\Gamma (s) = \int^\infty_0 t^{s-1} P^{s} e^{-tP} dt \nonumber
\ee
or
\be
P^{-s} =\frac{1}{\Gamma (s)} \int^\infty_0 t^{s-1} e^{-tP} dt \ .
\ee
Taking trace of both sides we get\footnote{Though the convergence
of the integral (\ref{gamma}) is guaranteed if $\Re (s) > 0$ now
one should be careful because taking trace over the infinite-dimensional space may introduce new divergences, see below.}
\be
\zeta_P (s) \equiv \Tr P^{-s} = \frac{1}{\Gamma (s)} \int^\infty_0 t^{s-1} \Tr e^{-tP} dt \ .
\ee
Now, let us analyze the poles of the underintegral expression in (\ref{main}), $\zeta_P (s) \phi (s)$.

First, let us show that $\phi (s)$ has poles at $s = 0,-1,-2,...$ . We know that $\phi (s)$ is regular when $\Re (s)>0$ (the fundamental strip). Now consider $\tilde{\phi}(s)$ defined by
\be
\tilde{\phi}(s) = \sum^{\infty}_{k=0}\frac{\chi^{(k)}(0)}{k!}\frac{1}{s+k}
\ee
and  consider the following integral
\be
\tilde{\chi} (p) = \frac{1}{2\pi i} \int^{c+i\infty}_{c-i\infty} p^{-s} \tilde{\phi} (s) ds \ ,
\ee
where we again assume that the contour could be closed to encircle all of poles. Then using the Cauchy residue theorem\footnote{Here by $\chi^{(k)}$ we denote in a standard way the $k$-th derivative of the cut-off function with respect to its argument $p$: $\chi^{(k)}=\frac{d^k\chi(p)}{dp^k}$.} we have
\be
\tilde{\chi} (p) = \frac{1}{2\pi i} \int^{c+i\infty}_{c-i\infty} p^{-s} \sum^{\infty}_{k=0}\frac{\chi^{(k)}(0)}{k!}\frac{1}{s+k} ds = \sum^{\infty}_{k=0}\frac{\chi^{(k)}(0)}{k!}p^k \equiv \chi (p)\ .
\ee
Hence, $\tilde{\phi}(s)\equiv\phi (s)$, which proves the above statement about the poles. The last equality in the above formula is valid within the analytic convergence region of the Maclaurin series of the cut-off function $\chi(p)$. In what follows, we will assume that the convergence radius is infinite.

Now, let us consider the poles of $\zeta_P (s)$. For this, we will use the following asymptotic small $t$ expansion  for the traced heat kernel coefficients:
\be\label{heat_kernel}
\Tr e^{-tP}\simeq \sum_{n\geqslant 0} t^{\frac{n-d}{m}}a_n (P) \ ,
\ee
where $d$ is the dimension of the manifold $\mathcal{M}$, $m$ is the order of $P$ and $a_n (P)$ are defined by
\be\label{a_n}
a_n (P) = \int_\mathcal{M} a_n (x,P)\sqrt{g}\, d^d x
\ee
for some known Seeley-DeWitt coefficients $a_n (x,P)$.

By the inverse Mellin transform (\ref{inverseMellin}), we can write for small $t$
\be
\Tr e^{-tP}\simeq \sum_{n\geqslant 0} t^{\frac{n-d}{m}}a_n (P) = \frac{1}{2\pi i} \int^{c+i\infty}_{c-i\infty} t^{-s} \Gamma (s) \zeta_P (s) ds \ .
\ee
Once again being sloppy about the contour and not pretending to be rigorous we can read of poles of $\Gamma (s) \zeta_P (s)$:
\be
{\rm Res} (\Gamma (s) \zeta_P (s))|_{s=\frac{d-n}{m}} = a_n (P) \ .\label{poles}
\ee

Let us now specify to the case $m=2$ and $d=4$. Then $a_n (P) =0$ for all odd $n$'s.\footnote{This is true for the case of the manifold without a boundary. When the boundary is not trivial, one also has $a_n (P) \ne 0$ for odd $n$, see e.g. \cite{Fursaev:2011zz}.\label{footnote}} Then we see from (\ref{poles}) for $n=4,6,8,...$ ($s=0,-1,-2,...$) that all poles come from the $\Gamma$-function and $\zeta_P (s)$ is regular and equal
\be
\zeta_P (s)= \frac{1}{{\rm Res} (\Gamma (s))|_{s=\frac{d-n}{m}} } a_n (P)\equiv (-1)^s s!|_{s=|\frac{d-n}{m}|} a_n (P)\ .
\ee
On the other hand, when $n=0,2$ ($s=1,2$) the poles should come from $\zeta_P (s)$ because $\Gamma(s)$ is regular:
\be
{\rm Res}\, \zeta_P (s)|_{s=1,2}=a_{0,2} (P)\ .
\ee

Now, we can evaluate (\ref{main}) using the information about the poles of $\zeta_P (s) \phi (s)$ from the previous paragraph.
\bea\label{general_trace}
& &\Tr\chi (P) = \frac{1}{2\pi i} \int^{c+i\infty}_{c-i\infty} \zeta_P (s) \phi (s) ds = \nonumber\\
&=& \phi(2) a_0 (P) + \phi(1) a_2 (P) + \sum_{s=0}^{\infty}(-1)^s \chi^{(s)}(0) a_{2(s+2)}(P)\equiv \sum_{k=0}^{\infty}f_{2k} a_{2k}(P)\,,
\eea
where (using the definition of $\phi(s)$ as a Mellin transform of $\chi (p)$)
\be
f_0 = \phi(2)\equiv \!\int^\infty_0\! p \chi (p)\,dp\ ,\ f_2 = \phi(1)\equiv \!\int^\infty_0\! \chi (p)\,dp\ , \ f_{2(2+k)}=(-1)^k \chi^{(k)}(0)\ ,k\geqslant 0\,.
\ee

\pdfbookmark[1]{D General equations of motion}{name13}
\hypersetup{bookmarksdepth=-2}
\section{General equations of motion}\label{AppEOM}
\hypersetup{bookmarksdepth}

The simplest way to calculate the variation of the Riemann tensor and all of its contractions is to vary directly the defining formulas \eqref{sriem}, \eqref{sric} and \eqref{signrs}, taking into account that the variation of the Christoffel symbols $\Gamma^\sigma{}_{\nu\rho}$ are tensors given by
\begin{eqnarray}\label{varCrist}
\delta\Gamma^{\sigma}{}_{\nu\rho}=\frac{1}{2}g^{\sigma\kappa}\left( \nabla_\nu h_{\rho\kappa} + \nabla_\rho h_{\nu\kappa} - \nabla_\kappa h_{\nu\rho} \right) =\frac{1}{2}\left( \nabla_\nu h_{\rho}{}^\sigma + \nabla_\rho h_{\nu}{}^\sigma - \nabla^\sigma h_{\nu\rho} \right) =: {\cal C}^{\sigma}{}_{\nu\rho}\ ,
\end{eqnarray}
where $h_{\mu\nu}:=\delta g_{\mu\nu}$. Then one trivially gets the linear part of the variation:
\begin{eqnarray}\label{varRiem}
\delta R_{\mu\nu\rho}{}^{\sigma} = 2 \nabla_{[\mu}{\cal C}^{\sigma}{}_{\nu ]\rho}=-g^{\sigma\kappa}\left( R_{\mu\nu (\rho}{}^{\lambda}h_{\kappa )\lambda} - 2\nabla_{[\mu}\nabla_{[\![\rho} h_{\nu ]\kappa ]\!]} \right),
\end{eqnarray}
where we used (\ref{Commut. Cov. Derv.Gen. Tensor.}) and in all the formulas of this appendix we use anti-symmetrization exclusively inside pairs of indices (here never between three or more indices) and we denote it by brackets $[\cdots]$ or $[\![\cdots]\!]$ as explained in the appendix \ref{apA}.  The variations of the other geometric objects trivially follow from (\ref{varRiem}). For example, the variations relevant for getting the equations of motion (\ref{Ricci flat EOM}) from the action in (\ref{grav action R linear}) are
\begin{eqnarray}\label{varR}
&& \delta R_{\mu\nu\rho\sigma} =R_{\mu\nu[\rho}{}^{\kappa}h_{\sigma]\kappa}+2\nabla_{[\mu}\nabla_{[\![\rho}h_{\sigma]\!]\nu]} \ ,\nn\\
&& \delta R^{\mu\nu}{}_{\rho\sigma} = - R^{[\mu \kappa}{}_{\rho\sigma}h^{\nu]}{}_{\kappa} + 2 \nabla_{[\![\rho}
\nabla^{[\mu}h_{\sigma ]\!]}{}^{\nu ]}\ ,\nn\\
&& \delta R^{\mu\nu\rho\sigma} = -2 R^{\rho\sigma[\mu}{}_\kappa h^{\nu]\kappa} -R^{[\rho\kappa\mu\nu}h^{\sigma]}{}_\kappa+2\nabla^{[\mu}\nabla^{[\![\rho} h^{\sigma]\!]\nu ]} \ ,\nn\\
&& \delta R_{\mu\nu} =\nabla^{\kappa}\nabla_{(\mu}h_{\nu)\kappa} - \frac{1}{2}\nabla_{\mu}\nabla_{\nu} h - \frac{1}{2}\square h_{\mu\nu}\ ,\nn\\
&& \delta R^{\mu\nu} =-2 R^{(\mu\kappa}h^{\nu)}{}_\kappa +\nabla^{\kappa}\nabla^{(\mu}h^{\nu)}{}_{\kappa} - \frac{1}{2}\nabla^{\mu}\nabla^{\nu} h -\frac{1}{2}\square h^{\mu\nu}\ ,\nn\\
&& \delta R = -R^{\mu\nu}h_{\mu\nu} +  \nabla^{\mu}\nabla^{\nu}h_{\mu\nu} - \Box h\ .
\end{eqnarray}
To vary the Weyl--dominated form of the action (\ref{grav action W}), one also needs the variation of the Weyl tensor in $d=4$ space (or spacetime) dimensions:
\begin{eqnarray}\label{varW}
\delta C_{\mu\nu\rho\sigma} &=&  C_{\mu\nu [\rho\kappa}h_{\sigma ]}{}^{\kappa} + R_{[\mu\kappa}g_{\nu ][\![\rho}h_{\sigma ]\!]}{}^{\kappa} + 2\nabla_{[ \mu}\nabla_{[\![\rho}h_{\sigma]\!]\nu ]}  + \nn\\
&+& 2g_{[\mu[\![\rho}\left( \nabla^{\kappa}\nabla_{(\sigma]\!]}h_{\nu])\kappa} - \frac{1}{2}\nabla_{\sigma]\!]}\nabla_{\nu]} h - \frac{1}{2}\square h_{\sigma]\!]\nu]} \right) + \nn\\
&+&  R_{[\mu[\![\rho}h_{\sigma]\!]\nu]} - \frac{1}{3}R g_{[\mu[\![\rho}h_{\sigma]\!]\nu]} + \frac{1}{3}g_{\mu[\rho}g_{\sigma]\nu} \left( R^{\kappa\lambda}h_{\kappa\lambda} -\nabla^{\kappa}\nabla^{\lambda}h_{\kappa\lambda} + \Box h \right).
\end{eqnarray}

Using these variations (and the related ones) the general form of the tensor of classical equations of motion, $E^{\alpha\beta}$, resulting from the action in \eqref{grav action R}, in the Riemann basis is given by (to be symmetrized with respect to $(\alpha,\beta)$ pair of indices, if needed):
\begin{eqnarray}\label{EOM general}
&&E^{\alpha\beta}=\frac{\Lambda^4\mu_0}{2}g^{\alpha\beta}+ \Lambda^2 \mu_1 \left( R^{\alpha\beta}-\frac{1}{2}g^{\alpha\beta}R \right) - \mu_2\left( 2R^{\alpha\beta}R+\frac{3}{2}g^{\alpha\beta}R^{\mu\nu}R_{\mu\nu}- \right.\nn\\ &&\left. -\frac{1}{2}g^{\alpha\beta}R^{2}+6R^{\mu\nu}R^{\alpha}{}_{\mu}{}^{\beta}{}_{\nu}\right) +\frac{\mu_3}{\Lambda^2}\left( \frac{504}{5}R^{\alpha\beta}{}^{;\mu}{}_{\mu}-\frac{84}{5}g^{\alpha\beta}R{}^{;\mu}{}_{\mu}- \frac{168}{5}R{}^{;\alpha\beta}+ \right.\nn\\ &&+\frac{13}{15}g^{\alpha\beta}R^{\mu\nu}R_{\mu}{}^{\rho}R_{\nu\rho}-\frac{13}{2}R^{\alpha\beta}R^{\mu\nu}R_{\mu\nu}-2R^{\alpha\mu}R^{\beta\nu}R_{\mu\nu}+ \frac{13}{4}g^{\alpha\beta}R^{\mu\nu}R_{\mu\nu}R-\nn\\ &&-\frac{4}{5}R^{\alpha\mu}R^{\beta}{}_{\mu}R+\frac{27}{10}R^{\alpha\beta}R^{2}-\frac{9}{20}g^{\alpha\beta}R^{3}+ \frac{63}{10}g^{\alpha\beta}R^{\mu\nu}R^{\rho\sigma}R_{\mu\rho\nu\sigma}+\frac{1}{5}R^{\alpha\beta}R^{\mu\nu\rho\sigma}R_{\mu\nu\rho\sigma}-\nn\\ &&-\frac{1}{10}g^{\alpha\beta}RR^{\mu\nu\rho\sigma}R_{\mu\nu\rho\sigma}-\frac{1}{30}g^{\alpha\beta}R^{\mu\nu\tau\omega}R_{\mu\nu}{}^{\rho\sigma}R_{\rho\sigma\tau\omega}+ \frac{61}{5}RR^{\alpha\mu\beta\nu}R_{\mu\nu}-4R^{\alpha\nu\mu\rho}R^{\beta}{}_{\mu}R_{\nu\rho}-\nn\\ &&-\frac{73}{5}R^{\alpha\nu\beta\rho}R^{\mu}{}_{\rho}R_{\mu\nu}+16R^{\alpha\rho\beta\sigma}R^{\mu\nu}R_{\mu\rho\nu\sigma}+ \frac{2}{5}RR^{\alpha\mu\nu\rho}R^{\beta}{}_{\mu\nu\rho}+\frac{1}{5}R^{\alpha\mu\nu\rho}R^{\beta}{}_{\mu}{}^{\sigma\tau}R_{\nu\rho\sigma\tau}-\nn\\ &&-\frac{2}{5}R^{\alpha\mu\rho\sigma}R^{\beta\nu}{}_{\rho\sigma}R_{\mu\nu}-\frac{2}{5}R^{\alpha\mu\nu\rho}R^{\beta}{}_{\nu}{}^{\sigma\tau}R_{\mu\rho\sigma\tau}- \frac{4}{5}R^{\alpha\mu\nu\rho}R^{\beta\sigma}{}_{\nu}{}^{\tau}R_{\mu\sigma\rho\tau}-\frac{57}{10}RR^{\alpha\beta}{}^{;\mu}{}_{\mu}-\nn\\ &&-\frac{43}{10}R^{\alpha\beta}R{}^{;\mu}{}_{\mu}+\frac{43}{20}g^{\alpha\beta}RR{}^{;\mu}{}_{\mu}-\frac{23}{10}R^{;\alpha\mu}R^{\beta}{}_{\mu}- \frac{17}{5}R^{\alpha\beta;\mu}R_{;\mu}+\frac{41}{40}g^{\alpha\beta}R^{;\mu}R{}_{;\mu}+\nn\\ &&+8R^{\alpha\beta;\mu\nu}R_{\mu\nu}-\frac{1}{5}g^{\alpha\beta}R^{\mu\nu}R{}_{;\mu\nu}+\frac{43}{5}R^{\alpha\mu;\nu}{}_{\nu}R^{\beta}{}_{\mu}+ 3R^{\alpha\beta;\mu}{}_{\mu}{}^{\nu}{}_{\nu}-\frac{3}{10}g^{\alpha\beta}R{}^{;\mu}{}_{\mu}{}^{\nu}{}_{\nu}-\nn\\ &&-\frac{5}{36}R^{\alpha\mu;\nu}R^{\beta}{}_{\nu;\mu}+\frac{39}{5}R^{\alpha\mu;\nu}R^{\beta}{}_{\mu;\nu}-\frac{4}{5}R^{\alpha\mu\beta\nu}R_{;\mu\nu}- 8g^{\alpha\beta}R^{\mu\nu}R_{\mu\nu}{}^{;\rho}{}_{\rho}-14R^{\alpha\mu\beta\nu}R_{\mu\nu}{}^{;\rho}{}_{\rho}-\nn\\ &&-8R^{\alpha\mu\beta\nu;\rho}{}_{\rho}R_{\mu\nu}+\frac{3}{10}g^{\alpha\beta}R^{\mu\rho;\nu}R_{\mu\nu;\rho}- \frac{9}{2}g^{\alpha\beta}R^{\mu\nu;\rho}R_{\mu\nu;\rho}-16R^{\alpha\mu\beta\nu;\rho}R_{\mu\nu;\rho}+\frac{8}{5}R^{\alpha\rho\mu\nu}R^{\beta}{}_{\mu;\nu\rho}-\nn\\ &&-\frac{2}{5}R^{\alpha\mu\nu\rho;\sigma}R^{\beta}{}_{\sigma\nu\rho;\mu}-2g^{\alpha\beta}R^{\mu\rho\nu\sigma}R_{\mu\nu;\rho\sigma}+ \frac{2}{5}g^{\alpha\beta}R^{\mu\nu\rho\sigma}R_{\mu\nu\rho\sigma}{}^{;\tau}{}_{\tau}+\frac{2}{5}g^{\alpha\beta}R^{\mu\nu\rho\sigma;\tau}R_{\mu\nu\rho\sigma;\tau}+\nn\\ &&+\frac{7}{5}R^{\mu\nu;\alpha}R^{\beta}{}_{\mu;\nu}-\frac{19}{10}R^{\alpha\mu;\beta}R_{;\mu}-4R^{\alpha\mu\nu\rho;\beta}R_{\mu\nu;\rho}- \frac{53}{5}R^{\alpha\mu;\nu\beta}R_{\mu\nu}-4R^{\mu\nu;\rho\alpha}R^{\beta}{}_{\mu\nu\rho}-\nn\\ &&\left. -\frac{2}{5}R^{\mu\nu\rho\sigma}{}^{;\alpha}R_{\mu\nu\rho\sigma}{}^{;\beta}+3R^{\mu\nu;\alpha\beta}R_{\mu\nu}+\frac{7}{10}RR^{;\alpha\beta}- \frac{2}{5}R^{\mu\nu\rho\sigma;\alpha\beta}R_{\mu\nu\rho\sigma}-\frac{6}{5}R{}^{;\mu}{}_{\mu}{}^{\alpha\beta}\right)\ .
\end{eqnarray}
This result is independent of the signature since it is written in a generally covariant way. Therefore, it holds both for Euclidean spaces and also for Lorentzian spacetimes of any dimension.

\pdfbookmark[1]{Acknowledgements}{name14}
\section*{Acknowledgements}
The authors would like to thank Ilya Shapiro for the discussions during different stages of this project. The work of LR was partially supported by the CAPES/PNPD post-doctoral grant and RM was supported by CNPq Masters and by CAPES PhD fellowships.

\newpage
\pdfbookmark[1]{References}{name15}
\bibliographystyle{utphys}
\bibliography{actiondiracbib}

\end{document}